# Effect of Transverse Gust on Free-Falling Plates

Jawahar Sivabharathy Samuthira Pandi, Ahmet Gungor, Chandan Bose, Antonio Attili, Ignazio Maria Viola

**Abstract**

The effects of transverse gusts on free-falling plates are investigated using two-way coupled fluid-structure interaction simulations for a Galilei number ($Ga$) between 10 and 50 and a density ratio ($\rho$) between 5 and 50. We consider a gust ratio ($G_R$) up to 5, where $G_R$ is the ratio of the free-stream velocity change to an estimate of the terminal velocity. We demonstrate that the plate experiences the gust as a transient horizontal force, which displaces it horizontally. This results in a transient reduction in the angle of attack, an increase in absolute velocity, and the generation of circulation. The vertical component of the latter increases the upward aerodynamic force, slowing down the vertical descent of the plate. Furthermore, the plate's horizontal displacement with respect to its original wake results in a further transient increase in the upward aerodynamic force. The altitude gained by the plate in response to the gust is maximum for $\rho = 15$, and increases non-monotonically with $Ga$ and $G_R$. The non-monotonic trend is due to plate pitch: if the maximum pitch of the plate in response to the gust is close to vertical, the plate temporarily falls faster, losing some of the altitude it has gained. The present findings reveal an energy-harvesting mechanism that free-falling bodies can exploit to increase their time afloat.

## 1 Introduction

The dynamics of solid bodies settling under gravity have garnered long-standing research interest due to their multi-fold applications in plant biology (Cummins *et al.*, 2018), meteorology (Kajikawa, 1982), and engineering (Marchildon *et al.*, 1964; Viets & Lee, 1971). When falling in a quiescent flow, or in a uniform and constant flow stream, the dynamics of discs or plates can be described by three non-dimensional groups, for instance: the length-to-thickness ratio $\chi$, the Reynolds number $Re$, and the dimensionless moment of inertia $I^*$ (Willmarth *et al.*, 1964; Smith, 1971). The kinematics of discs and plates have been studied experimentally (Willmarth *et al.*, 1964; Smith, 1971; Field *et al.*, 1997; Mahadevan *et al.*, 1999; Heisinger *et al.*, 2014; Tinklenberg *et al.*, 2023) and numerically (Auguste *et al.*, 2013; Jones & Shelley, 2005; Andersen *et al.*, 2005*a*; Lau *et al.*, 2018) over a wide range of $\chi$, $Re$, and $I^*$. These regimes are commonly mapped in a $Re - I^*$ phase diagram. Plates and disks fall stably for $Re$ below a critical threshold of the order of 100. Instead, above this threshold, their unsteady kinematics span from fluttering to tumbling with increasing $I^*$.(Field *et al.*, 1997; Andersen *et al.*, 2005*a*).

Quasi-steady force models have been used to study the dynamics of free-falling plates (Pesavento & Wang, 2004; Andersen *et al.*, 2005*a*,*b*) and flapping insect wings (Ellington, 1984; Sane, 2003). Fluid loads are decomposed into gravitational, buoyancy, quasi-steady circulatory lift and drag, and added mass. These models can predict complex plate motions, including fluttering and tumbling (Andersen *et al.*, 2005*b*). They have been extended to plates with inhomogeneous mass distribution (Huang *et al.*, 2013; Li *et al.*, 2022), flexible plates (Tam *et al.*, 2010), elongated cylinders (Letessier *et al.*, 2025), and flapping wings with finite aspect ratio (Wang *et al.*, 2016; Lee *et al.*, 2016). However, these models do not account for the history effect and thus the unsteady variations of lift and drag, including those due to non-fully developed wakes, leading-edge vortex formation and shedding, etc.

In contrast to bodies free-falling in a quiescent flow, the dynamics of free-falling bodies in a spatially or temporally varying flow stream are far less understood. The effects of these disturbances, or *gusts*, have been investigated for aircraft (Hunsaker & Wilson, 1915; Kussner, 1932; Wu *et al.*, 2019; Fidkowski *et al.*, 2008), and more recently for low-inertia flyers such as nano- and micro-aerial vehicles (Mohamed *et al.*, 2016; Acar *et al.*, 2020; Andreu-Angulo *et al.*, 2025). Gusts can disrupt their trajectory, orientation, and stability. While gusts are a hazard to aerial vehicles, they are essential for the long-distance dispersal of plant seeds (Horn *et al.*, 2001; Nathan *et al.*, 2011). Diaspores of plants such as the dandelion (*Taraxacum officinale*), the fleabane (*Conyza bonariensis*), and the common milkweed (*Asclepias syriaca*) travel vast distances relying only on wind power (Sacchi, 1987; Savage *et al.*, 2014; Cummins *et al.*, 2018; Galler & Rival, 2021; Viola & Nakayama, 2022). This biological observation has inspired the development of bio-inspired insect-scale flying sensors that are passively transported by the wind (Kim *et al.*, 2021; Iyer *et al.*, 2022; Chen *et al.*, 2023; Yang *et al.*, 2024). Therefore, a comprehensive understanding of gust interactions is crucial for informing the design of these ultra-light aerial systems (Regan & Jutte, 2012; Iyer *et al.*, 2022).

Gust encounters are classified into continuous and discrete gusts. The former are periodic or chaotic fluctuations and can be caused, for example, by atmospheric turbulence (Solari, 1987; Leahy, 2008). They might have a significant impact on the kinematics and wake dynamics of settling particles (Catalán *et al.*, 2024). Conversely, discrete gusts are spatially and temporally isolated disturbances. This study focuses on discrete gusts, for which a comprehensive review is provided in Jones *et al.* (2022). Gust encounters are further categorised based on the spatial and temporal evolution of the velocity field with respect to a body-fixed frame. A vortex gust is the disturbance due to a free vortex in a uniform stream. The effect of the vortex gust depends on the distance between the body and vortex trajectory, and whether the vortex impinges on the body (Muscutt *et al.*, 2017; Gungor *et al.*, 2022). It also depends on the vortex's circulation and orientation (Lagopoulos *et al.*, 2020; Pourfarzan & Wong, 2022; Gungor *et al.*, 2024*a*,*b*). Streamwise gusts are temporal variations of the magnitude of the relative velocity between the body and the free stream (Wang *et al.*, 2024). Finally, transverse gusts are flow disturbances orthogonal to the initial relative fluid-body velocity (Perrotta & Jones, 2017; Corkery *et al.*, 2018; Leung *et al.*, 2018; Biler *et al.*, 2019; Andreu-Angulo *et al.*, 2020; Sedky *et al.*, 2020; Grubb *et al.*, 2020; Olson *et al.*, 2021). For example, a body experiences a transverse gust when falling vertically in a stream that is spatially or temporally varying in the horizontal direction. This study will focus on a temporally varying horizontal gust.

The strength of streamwise and transverse discrete gusts is typically measured by the gust ratio $G_\mathrm{R}$ (Jones *et al.*, 2022). This is the ratio between the amplitude of the streamwise or transverse velocity variation and the initial relative fluid-body velocity. Several canonical functional forms of velocity variations have been considered, including steps, linear ramps, and sinusoidal functions (Wu *et al.*, 2019; Badrya *et al.*, 2022). Between the latter, this study employs the one-minus-cosine model, also known as the Pratt profile (Pratt, 1953; Lone & Dussart, 2019).

Linear potential-flow models provide a foundational framework for predicting the unsteady aerodynamic response of thin plates to transverse gusts. Wagner (1925) investigated the problem of a plate subjected to a temporal step change in transverse velocity. Theodorsen (1949) considered a translating plate in a sinusoidally time-varying transverse velocity field. Küssner (1936) considered a plate that translates through a constant stream featuring a spatial step change in the transverse velocity, such that the plate experiences the gust progressively from the leading to the trailing edge. Sears (1941) considered a plate translating through a constant but spatially varying sinusoidal transverse velocity field. These models assume that the perturbation in velocity is small and neglect the body's dynamic response to load fluctuations. Therefore, they are not applicable to low-inertia free-falling bodies and large gust ratios (Perrotta & Jones, 2017; Ōtomo *et al.*, 2021).

Transverse gusts have been experimentally investigated using various approaches in wind and water tunnels. For example, a body in a uniform and constant stream can be plunged transversely to the stream (Arredondo-Galeana *et al.*, 2021). Alternatively, a body can be simultaneously towed and plunged in a still fluid (Perrotta & Jones, 2018). Others have dynamically bent the streamlines of a fixed body in a wind tunnel by deflecting the side walls of the test section (Holmes, 1973, April; Fernandez *et al.*, 2021).These methods are not suitable for testing untethered bodies. Conversely, Viola *et al.* (2025) have developed an accelerating wind tunnel, where an upward, uniform, constant airstream allows the body to hover in the test section. Then, the entire wind tunnel is horizontally accelerated to simulate a horizontal gust, allowing the dynamic response of the untethered body to be studied. A similar approach is adopted in this numerical study, which focuses on the dynamic response of two-dimensional plates. This is the first study on the dynamic response of free-falling plates to transverse gusts.

Previous research on tethered bodies shows that a discrete transverse gust typically results in a sharp lift peak, often followed by an opposite-sign peak, and by a slow recovery to the pre-gust state (Perrotta & Jones, 2018; Moriche *et al.*, 2021*b*). The peak lift typically increases monotonically with the gust ratio (Jones & Cetiner, 2021; Moriche *et al.*, 2019) and coincides with the formation of a leading-edge vortex (Moriche *et al.*, 2021*a*). For a given gust ratio, flat plates and aerofoils exhibit functionally similar lift responses while their drag differs considerably (Moriche *et al.*, 2021*b*). How the above translates in the case of untethered bodies is not known.

Specifically, this study aims to: (i) assess the effects of transverse gusts on the trajectory, descent velocity, and stability of free-falling plates, (ii) elucidate the mechanisms underlying the gust response, and (iii) characterise the effect of the governing parameters, particularly the Galilei number, the density ratio, and the gust ratio.

The remainder of the paper is structured as follows. First, we present the methodology (§2), including the governing equations in the noninertial frame used in this study, the definition of the gust velocity, how we solve the equations, the computational domain and boundary conditions, and the verification validation of the results. In presenting the results, we first discuss how plates fall when dropped from rest in a quiescent flow (§3). This will aid the interpretation of the gust response in the following sections. Subsequently, we demonstrate the main results of this work - that plates are always uplifted by a transverse gust - and we explore the underlying mechanism (§4). Finally, we characterise the gust response of plates for a range of the governing parameters (§5). The key findings of the study are summarised in §6.

# 2 Methodology

We model a two-dimensional plate with length $\hat{l}$ and length-to-thickness ratio $\chi$, falling freely under the influence of gravitational acceleration $\hat{\boldsymbol{g}}$. The hat over the symbols is used to indicate dimensional quantities, and bold indicates vectors. We define the gravitational velocity as $\hat{\boldsymbol{u}}_g \equiv [(\hat{\boldsymbol{W}} - \hat{\boldsymbol{B}})/(\hat{\rho}_f \hat{l})]^{1/2}$, where $\hat{\boldsymbol{W}}$ and $\hat{\boldsymbol{B}}$ are the weight and buoyancy force per unit span of the plate, respectively, and $\hat{\rho}_f$ is the fluid density. However, because the physics depends only on the difference between gravitational and buoyancy force, we define a reduced gravity acceleration, $\hat{\boldsymbol{g}}^* \equiv (1 - \rho^{-1})\hat{\boldsymbol{g}}$, where $\rho$ is the solid-to-fluid density ratio, such that $\hat{\boldsymbol{W}} - \hat{\boldsymbol{B}} = \hat{m}\hat{\boldsymbol{g}}^*$, where $\hat{m}$ is the plate's mass. Therefore, the gravitational velocity is $\hat{\boldsymbol{u}}_g = [(\rho/\chi)\hat{\boldsymbol{g}}^* \hat{l}]^{1/2}$. In the following, all quantities are made non-dimensional using the fluid density $\hat{\rho}_f$, the plate length $\hat{l}$, and the magnitude of the gravitational velocity, $|\hat{\boldsymbol{u}}_g|$.

## 2.1 Governing Equations

In this section, we present the set of equations that we solved in the non-inertial frame of reference. To do so, we first present the equations in the inertial frame, and then explain how we modified this set of equations to account for the fictitious forces.

The flow around the plate is governed by the continuity and the incompressible Navier-Stokes equation, which, in nondimensional form, are

$$\nabla \cdot \boldsymbol{u} = 0, \tag{1}$$

$$\frac{\partial \boldsymbol{u}}{\partial t} + (\boldsymbol{u} \cdot \nabla)\boldsymbol{u} = -\nabla p + \frac{1}{Ga}\nabla^2 \boldsymbol{u}, \tag{2}$$

where $\boldsymbol{u} = (u, v)$ is the velocity vector for an arbitrary inertial frame of reference $\mathrm{P}(x, z)$; $t$ is time, $p \equiv p_{\mathrm{ph}} - g\eta$, is the physical pressure ($p_{\mathrm{ph}}$) net of the hydrostatic pressure ($g\eta$), with $\eta$ being a vertical coordinate in the direction of $\boldsymbol{g}$ with origin where the hydrostatic pressure is zero; $Ga \equiv |\hat{\boldsymbol{u}}_{\mathrm{g}}|\hat{l}/\hat{\nu}$ is the Galilei number, with $\hat{\nu}$ being the fluid kinematic viscosity. The readers might be used to seeing the Reynolds number ($Re$) based on the terminal velocity of the body instead of the Galilei number in equation 2. However, we prefer using the Galilei number, which is based on the gravitational velocity, because the latter is known a priori, while the terminal velocity is the outcome of the simulations.

The rigid body motion is governed by the two Euler equations of motion,

$$m\dot{\boldsymbol{u}}_{\mathrm{b}} = \boldsymbol{F} + \boldsymbol{\gamma}, \tag{3}$$

$$I_{\mathrm{o}}\dot{\boldsymbol{\omega}}_{\mathrm{b}} + \boldsymbol{\omega}_{\mathrm{b}} \times I_{\mathrm{o}}\boldsymbol{\omega}_{\mathrm{b}} = \boldsymbol{T}_{\mathrm{o}}, \tag{4}$$

where $m$ is the mass of the body; $\boldsymbol{u}_{\mathrm{b}}$ is the velocity of the centre of gravity of the solid body with respect to an inertial frame, e.g. $\mathrm{P}(x, z)$; $\boldsymbol{F}$ is the fluid force exerted on the body; $\boldsymbol{\gamma}$ is a unit vector in the direction of $\boldsymbol{g}$; $I_{\mathrm{o}}$ is the body inertia with respect to a non-inertial body-fixed frame centred at the centre of gravity; $\boldsymbol{\omega}_{\mathrm{b}}$ is the angular velocity of the solid body with respect to the inertial frame $\mathrm{P}(x, z)$; and $\boldsymbol{T}_{\mathrm{o}}$ is the fluid torque about the centre of gravity.

For a plate, $m = \rho/\chi$. We consider a plate with a uniformly distributed mass and, thus, the centre of mass coinciding with the centre of buoyancy, and $I_{\mathrm{o}} = (1/12)(\rho/\chi)(1 + 1/\chi^2)$ is the plate inertia with respect to its spanwise axis through its centre.

For a body falling in quiescent flow, a convenient inertial frame for equations 1 to 4 would be a frame moving downwards at the terminal velocity of the body. It would require a constant uniform inlet velocity equal in magnitude and opposite in sign to the terminal velocity. This boundary condition would ensure a finite displacement of the plate within the computational domain.

If the body is falling in a uniform, constant horizontal flow stream, the body would fall at the same vertical terminal velocity as in quiescent flow, but would also have a terminal horizontal velocity equal to the horizontal flow stream. Therefore, a convenient inertial frame that allows a finite displacement of the body within the computational domain is one that moves downwards at the terminal velocity of the body and horizontally at the free stream velocity.

In this study, we consider a plate falling in a horizontally accelerating flow stream. After the plate has reached its terminal velocity in quiescent flow, the flow is accelerated horizontally by a horizontal pressure gradient. As the plate density is different from the flow density, the acceleration of the plate in response to the pressure gradient is different to that of the fluid, and a relative horizontal velocity between the body and the surrounding fluid is established. The focus of this study is the physics underpinning the response of the plate to this horizontal pressure gradient.

We found it convenient and insightful to study physics in a non-inertial frame. Specifically, we use the gravitational velocity as an estimate of the terminal velocity. Therefore, we adopt a frame $\mathrm{P'}(x', z')$, which, with respect to an earth-fixed frame P, moves horizontally at the free stream velocity $u_{\mathrm{G}}$, namely the *gust* velocity. It also moves vertically at the body's gravitational velocity. The latter is $\boldsymbol{u}_g = (u_g, v_g) = (0, -1)$. Therefore, the velocity and acceleration of P' observed by P are $\boldsymbol{u}_{\mathrm{P'}} = (u_{\mathrm{G}}, -1)$ and $\dot{\boldsymbol{u}}_{\mathrm{P'}} = (\dot{u}_{\mathrm{G}}, 0)$, respectively.

A key advantage of solving the governing equations in the frame P' is that the boundary conditions are constant. In fact, the flow velocity $\boldsymbol{u}'$ observed by P' at a sufficient distance from the body is $\boldsymbol{u}' = (0, 1)$, which is constant. This approach is the same as the one demonstrated by Viola *et al.* (2025) experimentally.

To find a set of equations in this non-inertial frame, we substitute $\boldsymbol{u} = \boldsymbol{u}' + \boldsymbol{u}_{\mathrm{P'}}$, and $\dot{\boldsymbol{u}}_{\mathrm{b}} = \dot{\boldsymbol{u}}'_{\mathrm{b}} + \dot{\boldsymbol{u}}_{\mathrm{P'}}$ in equations 1 to 4, where $\dot{\boldsymbol{u}}'_{\mathrm{b}}$ is the acceleration of the body's centre of mass with respect to P'. Notably, because P' does not rotate with respect to P, the Coriolis acceleration, the Euler acceleration, and the centripetal acceleration are null, and the only fictitious acceleration is the linear acceleration $\dot{\boldsymbol{u}}_{\mathrm{P'}}$, which is the gust acceleration: $\dot{\boldsymbol{u}}_{\mathrm{P'}} = \dot{\boldsymbol{u}}_{\mathrm{G}}$.

Because $\boldsymbol{u}_{\mathrm{P'}}$ is uniform, its spatial gradients are null, and thus the functional form of equation 1 remains unchanged. For the same reason, $\boldsymbol{u}_{\mathrm{P'}}$ does not appear in the non-inertial Navier-Stokes equation. Conversely, $\dot{\boldsymbol{u}}_{\mathrm{G}}$ (i.e. $\dot{\boldsymbol{u}}_{\mathrm{P'}}$) appears in equation 2 because of the first term on the left-hand side - the temporal acceleration. Nonetheless, like the gravitational force, this fictitious force, being uniform, generates a uniform hydrostatic pressure gradient that does not change the flow field. Therefore, similarly to gravitational acceleration, fictitious acceleration also does not need to be included in the momentum balance, and the non-inertial Navier-Stokes equation takes the same functional form as equation 2. Overall, the two non-inertial fluid equations are

$$\nabla \cdot \boldsymbol{u}' = 0, \tag{5}$$

$$\frac{\partial \boldsymbol{u}'}{\partial t} + (\boldsymbol{u}' \cdot \nabla)\boldsymbol{u}' = -\nabla p' + \frac{1}{Ga}\nabla^2 \boldsymbol{u}', \tag{6}$$

where $p' \equiv p_{\mathrm{ph}} - g\eta + \dot{u}_{\mathrm{G}}\epsilon$, and $\epsilon$ is a horizontal coordinate with origin at the centre of the computational domain (Viola *et al.*, 2025). Because the fluid governing equations for a non-inertial frame that accelerates with the gust do not include any momentum source term (equations 5 and 6), and the external boundary conditions on the velocity and pressure are constant for this pair of partial differential equations, we infer that the flow field would remain uniform and constant during the gust if there was no solid body within the computational domain. The only way the flow can be perturbed is through a solid body immersed in the fluid domain. Therefore, the gust is perceived by the flow field because the body within it moves in response to the gust. This conclusion was shown experimentally by Viola *et al.* (2025).

The body dynamics is governed by the Euler's equations. In the linear Euler's equation (equation 3), because the fluid force $\boldsymbol{F}$ computed from equations 5 and 6 does not include the horizontal buoyancy force generated by the horizontal hydrostatic pressure gradient, the buoyancy force must be added to the non-inertial first Euler equation. We do so by adopting the same approach as for the gravity force, and we define a reduced gust acceleration, $\dot{\boldsymbol{u}}^*_{\mathrm{G}} \equiv (1 - \rho^{-1})\dot{\boldsymbol{u}}_{\mathrm{G}}$, such that $-m\dot{\boldsymbol{u}}^*_{\mathrm{G}}$ is the sum of the fictitious force $-m\dot{\boldsymbol{u}}_{\mathrm{G}}$ and of the buoyancy force due to the uniform, horizontal hydrostatic pressure gradient that it generates. Finally, because the non-inertial frame P' does not rotate, the body's angular velocity observed by P' is the same as that observed by P, $\omega'_{\mathrm{b}} = \omega_{\mathrm{b}}$. Therefore, the two non-inertial Euler's equations of motion are

$$m\dot{\boldsymbol{u}}'_{\mathrm{b}} = \boldsymbol{F} + \boldsymbol{\gamma} - m\dot{\boldsymbol{u}}^*_{\mathrm{G}}, \tag{7}$$

$$I_{\mathrm{o}}\dot{\boldsymbol{\omega}}_{\mathrm{b}} + \boldsymbol{\omega}_{\mathrm{b}} \times I_{\mathrm{o}}\boldsymbol{\omega}_{\mathrm{b}} = \boldsymbol{T}_{\mathrm{o}}. \tag{8}$$

In the set of fluid and structural equations in the non-inertial frame P' accelerating with the gust (equations 1 to 4), the gust acceleration enters only as a linear fictitious force in the first Euler's equation. The effect of the gust is a body force that acts on the plate and that is proportional to the flow acceleration. Therefore, we infer that the effect of the gust is not initially perceived by the body as a change in the fluid loads ($\boldsymbol{F}$ and $\boldsymbol{T}_{\mathrm{o}}$), but the body is initially displaced by the force $-m\dot{\boldsymbol{u}}^*_{\mathrm{G}}$. Once the body moves in response of this force, the flow field varies and the following dynamics of the body is governed by the fluid-structure interaction.

The last two terms on the right-hand side of equation 7 are the non-dimensional reduced gravitational and gust accelerations. The nondimensional weight is $mg = 1$. These two accelerations are both known a priori and prescribed as input to the simulation. Because the functional form of the inertial governing equations 1 to 4 is the same as that of the non-inertial equations 5 to 8, we solved the latter equations with an inertial solver. Specifically, we modified the input term for the gravity acceleration to include $\gamma$ along the vertical direction $z$, and $-m\dot{u}^*_{\mathrm{G}}$ along the horizontal direction $x$. The function of the reduced gust acceleration $\dot{u}^*_{\mathrm{G}}$ with time is given by its integrand in the next section (§2.2).

## 2.2 Gust velocity

We consider the plate falling in an initially quiescent fluid, which then accelerates horizontally with a cosine ramp to a constant horizontal velocity. The non-dimensional horizontal velocity with respect to an inertial frame is (figure 1a):

$$u_G = \begin{cases} 0, & t \le t_s, \\ [2(1-\rho^{-1})]^{-1}\left[1-\cos\left(\frac{\pi(t-t_s)}{t_G}\right)\right] G_R, & t_s < t \le t_s + t_G, \\ (1-\rho^{-1})^{-1} G_R, & t > t_s + t_G, \end{cases} \tag{9}$$

where $t_s$ is the start time of the transverse gust; $t_G$ is the gust duration; and $G_R$ is the maximum amplitude of the velocity change.

The ratio between the amplitude of the velocity disturbance and the initial fluid-body velocity is known in the gust literature as the *gust ratio* (Jones *et al.*, 2022). Therefore, we call $G_R$ the gust ratio because it is proportional to the ratio between the velocity disturbance and the gravitational velocity, which is an estimate of the undisturbed body velocity. The proportionality constant $(1-\rho^{-1})^{-1}$ ensures that the reduced gust acceleration accounts for the gust force net of the buoyancy force. Therefore, the reduced gust acceleration is independent of buoyancy. Specifically, the time derivative of equation 9 is $\dot{u}_G$, and $\dot{u}_G^* = (1-\rho^{-1})\dot{u}_G$ is the horizontal reduced gust acceleration used as input in equation 8.

In summary, the governing physics depends on the aspect ratio $\chi$, which enters in the boundary conditions of equations 5 and 6). Together with the density ratio $\rho$, $\chi$ also defines uniquely the mass $m$ and the inertia $I_o$ in equation 7 and 8. The Galilei number $Ga$ enters in equation 6), and the horizontal reduced gust acceleration $\dot{u}_G^*$ enters in equation 7). For our form of the gust velocity in equation 9, the gust acceleration is uniquely defined by $G_R$ and $t_G$. In this study we consider a plate with $\chi = 10$, $\rho \in \{5, 10, ..., 50\}$, $Ga \in \{10, 20, ..., 50\}$, $G_R \in \{0, 1, ..., 5\}$, and $t_G \in \{0.2, 0.5, 1, 2, 3, 4, 5\}$.

## 2.3 Computational Methodology

Equations 5 to 8 are solved with two-way coupled fluid-structure interaction (FSI) simulations using `OpenFOAM v2306` (Jasak *et al.*, 2007). The fluid Equations 5 to 6 are solved with a finite volume approach. To incorporate the plate movement, we adopt an overset mesh. The moving part of the grid follows the plate, thereby maintaining the resolution and quality of the mesh. We utilise the `cellVolumeWeight` interpolation scheme (Guerrero, 2006; Chandar & Gopalan, 2016) to transfer velocity and pressure across the boundaries of moving and refinement zones. The six-degree-of-freedom motion of the rigid body is solved using the `sixDoFRigidBodyMotion` solver available within OpenFOAM v2306. The simulations are carried out using the solver `overPimpleDyMFoam`, coupled with the dynamic mesh handling provided by `dynamicMotionSolverFvMesh`. The `sixDoFRigidBodyMotion` solver is modified to generate the effect of horizontal gusts. A partitioned weak coupling strategy-based six-degrees-of-freedom solver is used, where the fluid and solid solvers interact in a staggered manner. The structural solver uses a Newmark-beta time integration scheme. The spatial and temporal discretisations are second-order accurate. A preconditioned bi-conjugate gradient (`PBiCGStab`) iterative solver is used to solve for the pressure and cell-displacement. The preconditioning is implemented using the diagonal-based incomplete LU preconditioner (`DILU`) method. The coupled pressure-velocity equation is solved via a preconditioned smooth solver along with the symmetric Gauss-Seidel method for preconditioning. The absolute error tolerance criteria for pressure and velocity are set to $10^{-7}$.

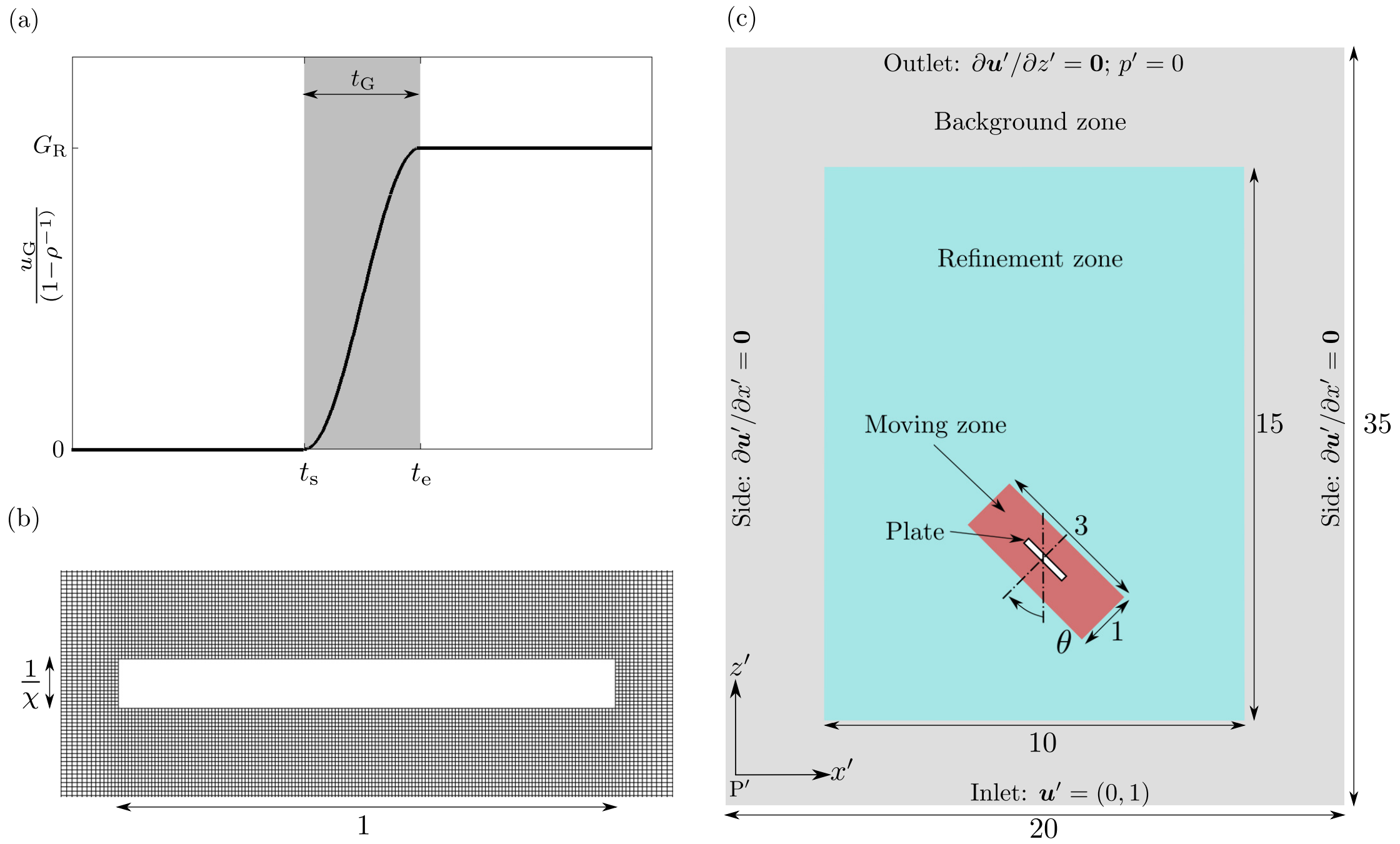

Figure 1: Gust response of the free-falling plate: (a) non-dimensional horizontal gust velocity with respect to an inertial frame, (b) close view of the mesh around the plate and (c) schematic of the computational domain (not-to-scale) and boundary conditions. The moving, refinement and background zones are highlighted.

## 2.4 Computational Domain

A schematic of the computational domain along with the boundary conditions is presented in figure 1a. A free-stream velocity $\boldsymbol{u}'$ in the vertical direction is prescribed at the inlet, with its magnitude matching that of the gravitational velocity. This is roughly equal to the expected terminal velocity of the plate, ensuring it remains within the refinement zone for the entire simulation. The velocity gradient normal to the boundary is set to zero on the two side boundaries and the outlet boundary. A uniform null pressure is prescribed on the outlet boundary. The mesh is made of three zones, namely the moving, refinement and background zones (figure 1a). The moving zone encompasses the plate boundary and exhibits both translational and rotational motions simultaneously. Instead, the refinement and background zones remain fixed. The mesh consists of $470,578$ hexahedral elements. A fine spatial resolution of the mesh is employed in the moving and refinement zones to resolve the boundary layer and the near-field wake (figure 1b). The moving zone and refinement zones consist of $25,578$ and $375,000$ elements, respectively. The grid spacing in the moving zone is non-uniform, with the height of the first grid on the plate surface being 0.007 times the length of the plate. The refinement and background zones have constant grid spacing of 0.02 and 0.1, respectively. The mesh resolution in the refinement zone is kept the same as that of the boundaries of the moving cell zone to minimise the interpolation error.

## 2.5 Verification and Validation

The numerical uncertainty is estimated for a plate with $\rho = 5$ and the highest $Ga$ ($Ga = 50$) and $G_{\mathrm{R}}$ ($G_{\mathrm{R}} = 5$). We found that the uncertainty at 95% confidence level due to the finite grid and time resolutions in the estimate of the altitude gain $Z_{\mathrm{G}}$, which is used in section 5.2 to quantify the gust response of the plate, is about 5%. Details are provided in Appendix A. As there are no experiments to validate the effect of gusts on free-falling plates, we validated the computed velocities with those of an aluminium plate with $\chi = 14$ while free-falling and fluttering in water (Andersen *et al.*, 2005*b*). The high length-to-span ratio, between 15 and 30, made the experiment quasi-two-dimensional. The Reynolds number in the experiments was calculated based on the average descent velocity over five fluttering periods across ten trials, yielding $Re = 1147$. The same grid and time step used in the rest of the study is used for this validation case. The numerical-experimental comparison shows a good agreement between the computed and measured horizontal, vertical and angular velocity of the plate with respect to an earth-fixed frame (figure 2).

# 3 Stability of free-falling flat plates in quiescent flow

First, the conditions governing the stability of flat plates falling under gravity are examined. We use the terms stable and unstable to describe the time-dependent characteristics of plate descent regimes. In the parameter space considered, two regimes are observed: steady fall, which is classified as stable, and fluttering motion, which is classified as unstable. Tumbling and chaotic motions are anticipated to occur only at higher values of $Ga$ and $\rho$ than those explored in this study (Smith, 1971; Field *et al.*, 1997; Andersen *et al.*, 2005*b*).

An initial disturbance is provided to test the stability of the kinematics for a range of $Ga$ and $\rho$. A velocity equal to half of the gravitational velocity is prescribed as the initial horizontal velocity. Specifically, $\boldsymbol{u}'_{\mathrm{b}} = (0.5, 1)$ at $t = 0$, while the initial orientation of the plate is horizontal ($\theta = 0$).

We observe that the initial horizontal velocity makes the plate flutter. After this initial transient, at low $Ga$ ($Ga \leq 30$), the vertical velocity fluctuation amplitude decreases exponentially, and eventually the plate falls steadily. Instead, at high Ga ($Ga \geq 40$), the vertical velocity fluctuation becomes periodic. This is shown, for example, for $\rho = 5$, by the Reynolds number $Re$ based on the instantaneous velocity with respect to an earth-fixed frame in figure 3a.

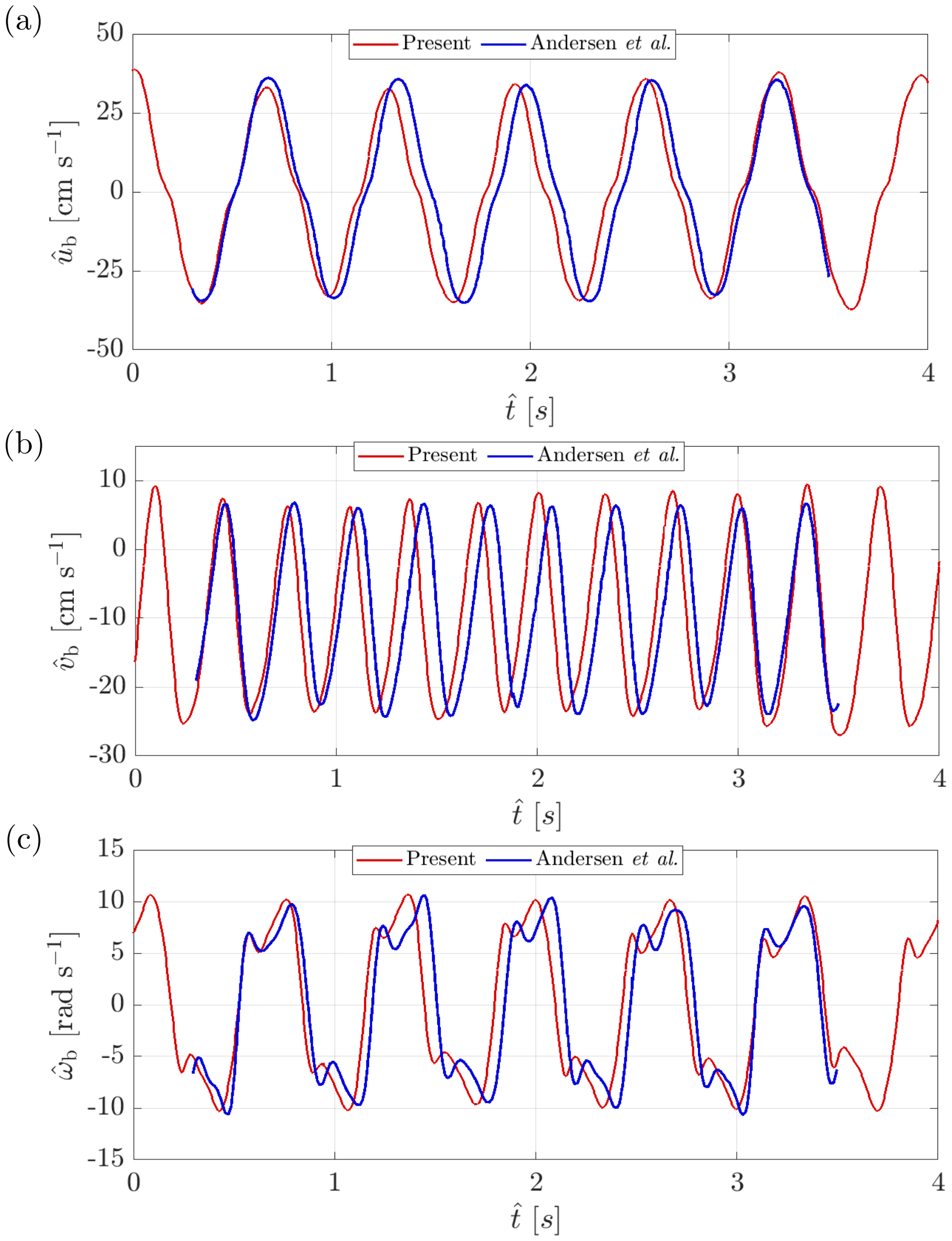

Figure 2: Time history of the (a) horizontal, (b) vertical, and (c) angular velocity of a fluttering plate with respect to an earth-fixed frame and comparison with the experiments of Andersen *et al.* (2005*b*). The results shown correspond to five fluttering cycles after the system has reached the periodic motion; therefore, $t = 0$ does not indicate the start of the simulations or experiments.

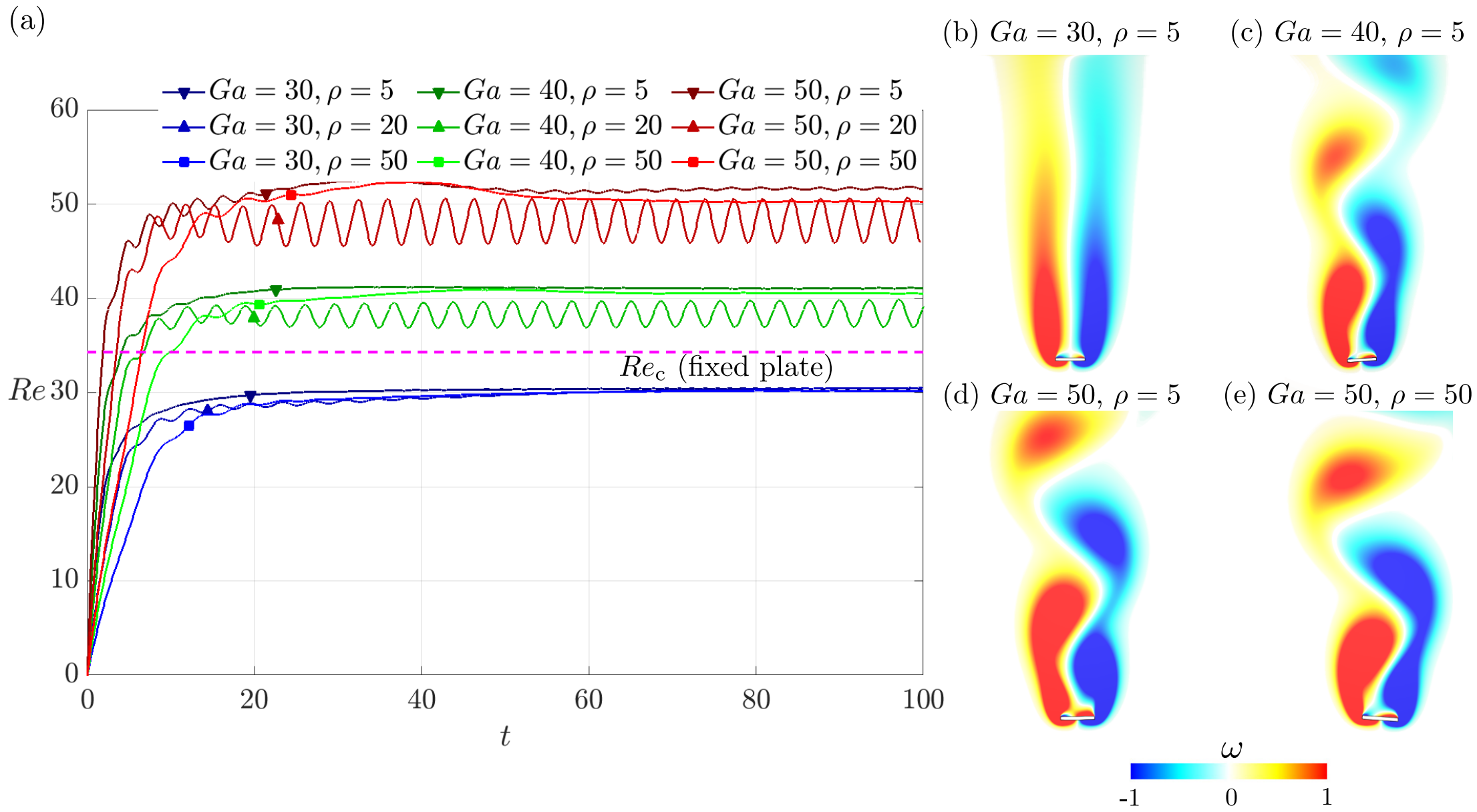

Figure 3: Free-falling plate in quiescent flow. (a) Time history of the Reynolds number ($Re$) based on the instantaneous descent velocity of the plate with respect to an earth-fixed frame and for various $Ga$ and $\rho$. The critical Reynolds number ($Re_c$) identifying the onset of unsteadiness of a fixed plate is marked by a broken line. Contours of the instantaneous vorticity $\omega$ for (b) $Ga = 30$ and $\rho = 5$, (c) $Ga = 40$ and $\rho = 5$, (f) $Ga = 50$ and $\rho = 5$, (e) $Ga = 50$ and $\rho = 50$.

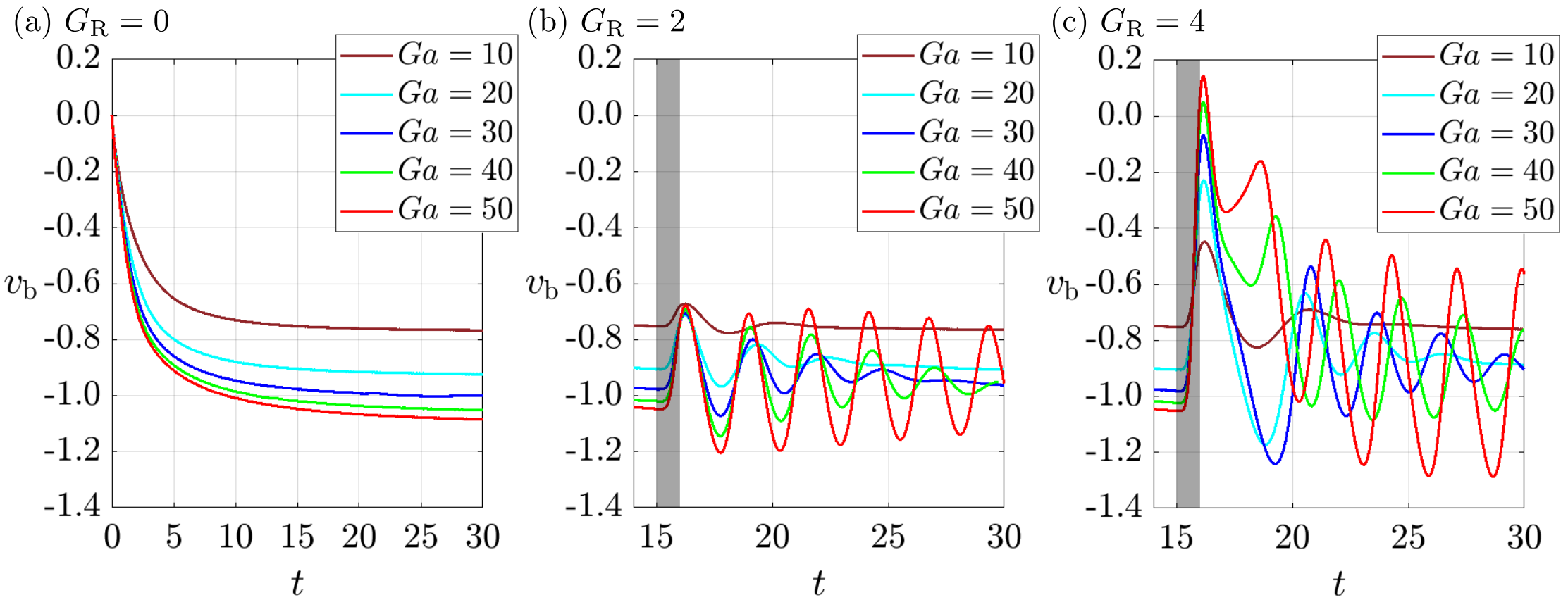

Figure 4: Time history of the plate's vertical velocity with respect to an earth-fixed frame for plates for different $Ga$ values, $\rho = 5$, and for (a) $G_R = 0$ (no gust); (b) $G_R = 2$; and (c) $G_R = 4$. The grey-shaded region shows the gust duration between $t = 15$ and 16.

The effect of $\rho$ is to modulate the amplitude of the oscillations of the kinematics. For example, at $t = 100$ and $\rho = 5$, the wake is stable and unstable at $Ga = 30$ and 40, respectively (figures 3b and c, respectively). At $Ga = 50$ and $\rho = 5$, the amplitude of the fluctuations is so small that it could be confused with a steady kinematics (figure 3a), but the wake is unstable (figure 3d), resulting in small pitch fluctuations. Therefore, the stability of the wake and of the kinematics depends exclusively on $Ga$.

The $Ga$ threshold between these two terminal regimes, either steady or fluttering, depends on whether the terminal velocity is smaller or greater than the critical Reynolds number $Re_c$ ($Re_c = 34.27$) at which the wake becomes unstable for a fixed plate. The critical Reynolds number ($Re_c$) identifying the onset of vortex shedding is computed using the Stuart-Landau equation (Thompson & Le Gal, 2004; Zhao *et al.*, 2014; Rastan & Alam, 2021). Details of this procedure are described in Appendix B.

Therefore, at any $\rho$ value, if $Re$ is greater than $Re_c$, vortex shedding occurs, resulting in flutter oscillations. The amplitude of these oscillations depends on $\rho$. Specifically, there exists a $\rho$ value that maximises the amplitude of the fluttering fluctuations. For example, for both $Ga = 40$ and 50, the amplitude of the vertical velocity fluctuations is maximum at $\rho = 20$ and lower at both $\rho = 5$ and 50 (figure 3a).

# 4 Uplifting mechanism of free-falling plates in transverse gusts

In this section, first, we show that the effect of a transverse gust on a free-falling plate is to decrease its descent velocity (§4.1). Therefore, in the rest of this section, we aim to explain the underlying mechanism. We achieve this by first demonstrating how the plate acquires a relative horizontal velocity with respect to the stream in response to the gust (§4.2). Then, we show that the plate rotation is not the main reason for the decrease in descent velocity (§4.3). Therefore, we consider the gust response of a plate that cannot rotate, and we use a quasi-steady force decomposition to reveal the underlying gust uplifting mechanism (§4.4). In this section, $\rho = 5$ and $t_G = 1$.

The results presented in this section are independent of the initial horizontal velocity of the free stream, while they only depend on the relative flow velocity between the free stream and the body. Therefore, results are presented for the body's horizontal coordinate $x'$ and velocity $u'_b$ observed by a frame P' moving horizontally at the free stream velocity and centred with the body before the gust. Conversely, results are presented for the vertical coordinate $z$ and velocity $v_{\mathrm{b}}$ observed by an Earth-fixed frame P, where zero altitude is where the plate is initially dropped from a stationary position.

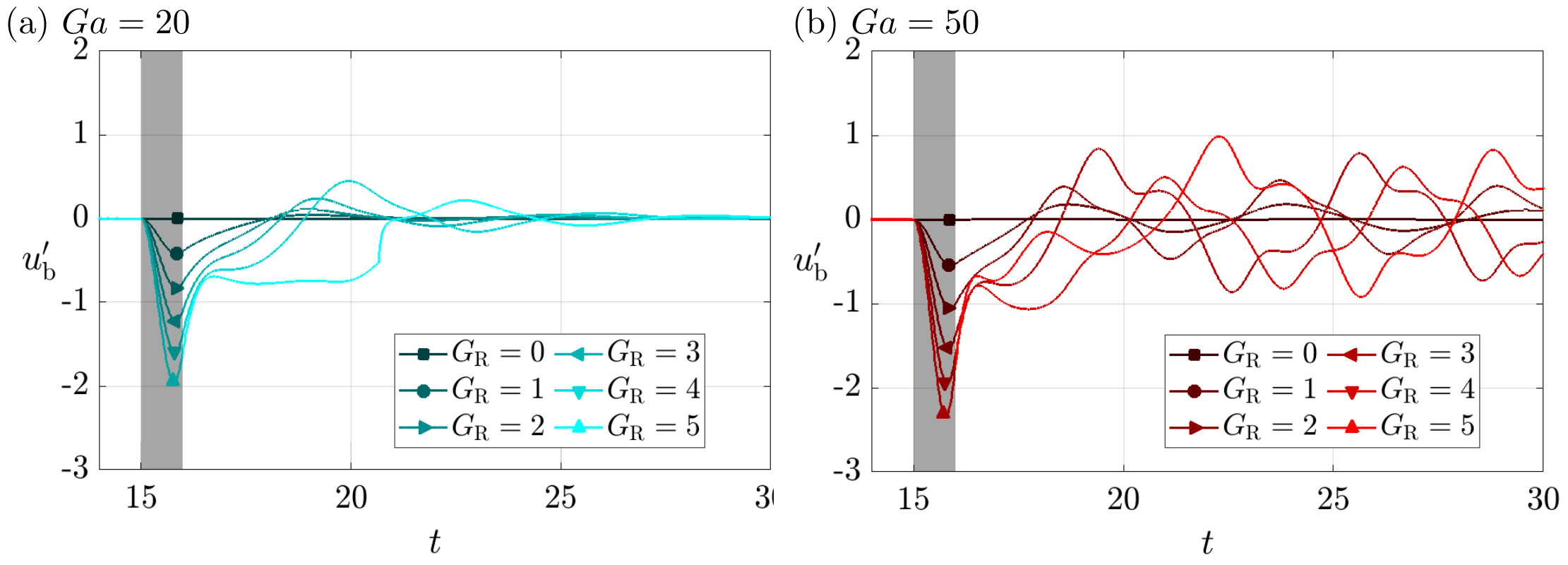

Figure 5: Time history of the horizontal plate velocity with respect to a non-inertial frame accelerating with the flow stream for $\rho = 5$ and (a) $Ga = 20$, and (b) $Ga = 50$. The grey-shaded region indicates the gust duration between $t = 15$ and $t = 16$.

## 4.1 Gust effect on descent velocity

We consider plates at $Ga \in \{10, 20, 30, 40, 50\}$. All plates are dropped from a stationary position and their vertical velocity $v_{\mathrm{b}}$ with respect to an earth-fixed frame tends asymptotically to a constant terminal velocity, which they reach at $t \approx 15$ (figure 4a). For example, for $Ga = 30$, the plate falls at $v_{\mathrm{b}} = 1$, and thus it falls at the gravitational velocity. Therefore, for $Ga = 30$, at the terminal velocity, $Re = 30$ (consistently with figure 4). For lower and higher $Ga$, the plate falls slower and faster than the gravitational velocity, respectively.

Until plates reach the terminal velocity, they show no or negligible velocity oscillations even at $Ga \geq 40$, for which $Re$ is greater than $Re_c$. A gust is applied between times $t = 15$ and 16 (shaded areas in figures 4b and 4c). Therefore, for every $Ga$, when the gust is applied, the plate falls steadily at the terminal velocity.

We discovered that, for every $Ga$, the effect of the gust is initially to slow down the descent velocity of the plate (figure 4b). This effect increases with increasing $Ga$ and gust ratio (figures 4b and 4c). For high $Ga$ and $G_{\mathrm{R}}$, the gust can also result in a temporarily positive vertical velocity ($G_{\mathrm{R}} = 4$ and $Ga = 40$ and 50 in figure 4c).

Following this initial reduction of the descent velocity, the plates oscillate and eventually reach either a stable kinematics or a periodic flattering kinematics, depending on whether their terminal velocity is lower or higher than $Re_c$, respectively. For example, the oscillations of the vertical velocity decrease exponentially for $Ga = 10$, 20 and 30, while they tend to a fixed amplitude for $Ga = 40$ and 50 (not shown in figure 4).

## 4.2 Gust effect on transverse velocity

Equation 6 revealed that the initial effect of the gust is that of a horizontal fictitious force $-m\dot{\boldsymbol{u}}^*_{\mathrm{G}}$. We also showed that the gust results in a temporary reduction in the descent velocity (section 4.1), which must be accompanied by an increase in the vertical force. Here, we explore how the *horizontal* fictitious force leads to a peak in the *vertical* upward force $F_z$.

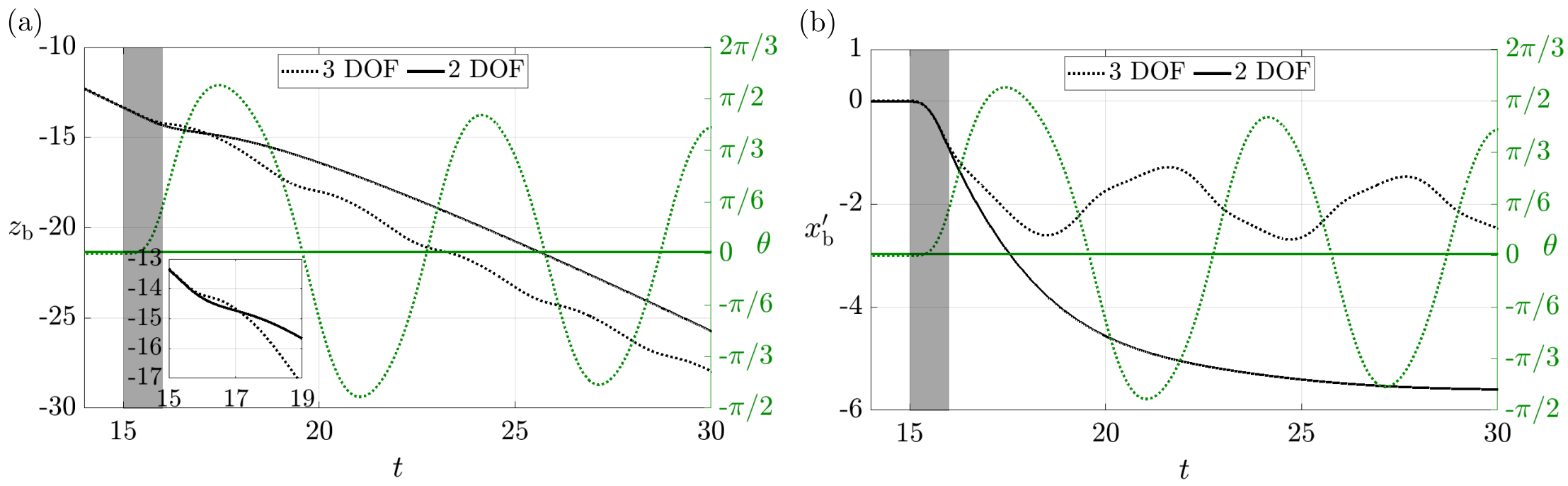

Figure 6: Time history of (a) the plate's altitude with respect to an earth-fixed frame and its pitch angle, and (b) the horizontal plate velocity with respect to a non-inertial frame accelerating with the flow stream and its pitch angle. Results are presented for $Ga = 50$, $\rho = 5$, and $G_R = 3$ for plates with two and three degrees of freedom. The grey-shaded region indicates the gust duration between $t = 15$ and $t = 16$.

Before the gust, the plate falls vertically at a constant terminal velocity. As discussed in section 2.1, in a uniform and constant horizontal flow stream, the plate moves horizontally at the stream velocity. Therefore, there is no relative velocity between the flow and the plate. When the stream accelerates, the plate also accelerates. However, if the body density is greater than that of the fluid ($\rho > 1$), it lags behind the flow and reaches the stream's new horizontal velocity after a finite time.

This behaviour can be clearly visualised using the frame P'$(x', z')$, which accelerates horizontally with the stream. In this frame, the plate's horizontal velocity is null both before and after the gust (figure 5a and 5b). If the stream accelerates, the plate accelerates in the opposite ($-x'$) direction with respect to the P' frame. The force displacing the plate in the $-x'$ direction is the fictitious force, $-m\dot{\boldsymbol{u}}^*_G$. We found that the negative peak in the horizontal velocity $u'_b$ increases monotonically with $G_R$, i.e. with the magnitude of the gust force $-m\dot{\boldsymbol{u}}^*_G$ (figure 5a and 5b).

Therefore, the horizontal velocity of the body with respect to P' establishes a relative horizontal velocity between the body and the fluid. Consequently, soon after the gust force decreases ($t = 15.5$), the horizontal velocity returns towards zero because of the fluid drag (figure 5a and 5b).

## 4.3 Free-to-rotate and rotationally constrained plates

As the plate gains horizontal velocity in response to the gust, it also pitches. In this section, we compare the gust response of a plate that can pitch with that of one whose pitch angle is fixed ($\theta = 0$). Specifically, we compare two plates at $Ga = 50$, $\rho = 5$, and $G_R = 3$ with $\mathbf{I}_o = \rho(1 + \chi^{-2})/(12\chi)$ and $\mathbf{I}_o \to \infty$, respectively. Therefore, the two plates have two and three Degrees of Freedom (DOF), respectively.

The time history of the vertical coordinate of the plate in an earth-fixed frame, $z_b$ (figure 6a), shows that, immediately after the gust (e.g. at $t = 16$), the plate with three DOF gains more altitude than that with two DOF (see inset in figure 6a). However, subsequently, the plate with three DOF loses its altitude when the pitch angle is close to $\pi/2$. Therefore, while the rotation of the plate helps it gain altitude, it eventually loses altitude because it falls faster when it reaches a vertical orientation.

The time history of the horizontal coordinate of the plate $u'_b$, with respect to a non-inertial frame that moves with the horizontal stream, shows that the constrained plate exhibits a higher horizontal range compared to that of the unconstrained case (figure 6b). After the end of the gust, the horizontal displacement of the plate reaches monotonically more than five diameters in the opposite direction of the gust. Conversely, the plate with finite-inertia begins to flutter in response to the gust around two diameters away from its original horizontal position.

We infer that the angular rotation of the plate is not essential to the uplifting mechanisms. It can enhance the initial uplift, but it also results in the early termination of the uplift. Furthermore, if the maximum pitch angle is close to vertical ($\pi(2n-1)/2$), the plate drops faster until it reaches a more horizontal orientation. This latter phenomenon is studied in section 5.3. Overall, for the cases tested, the constrained plate showed a higher gust uplift than the free-rotating plates.

## 4.4 Uplifting mechanism for rotationally-constrained plates

We demonstrated that the gust uplifting mechanism is caused by the generation of horizontal velocity, which results in a surge of vertical force (section 4.2). We clarified that the rotation of the plate is inessential to this mechanism (section 4.3). Here, we explore why the generation of horizontal velocity results in a vertical force surge for a rotationally-constrained plate (two DOF).

We break down the force components in quasi-steady lift and drag, added mass, and residual unsteady force. We write the equation of motion in non-dimensional form as:

$$\underbrace{\frac{\rho}{\chi}\dot{v}_\mathrm{b}+1}_{F_z} = \underbrace{\Gamma u_\mathrm{b}' u_\mathrm{rel}}_{F_z^\mathrm{C}} \underbrace{-\frac{1}{2}\left(C_\mathrm{D}^0\cos^2\alpha + C_\mathrm{D}^{\pi/2}\sin^2\alpha\right)u_\mathrm{rel}v_\mathrm{b}}_{F_z^\mathrm{D}} \underbrace{-\,m_\mathrm{a}\dot{v}_\mathrm{b}}_{F_z^\mathrm{AM}} + F_z^\mathrm{UW}. \tag{10}$$

The term on the left-hand side of equation 10, $F_z$, is the total vertical body force acting on the plate, where $\chi = 10$ for the plate. This term encompasses both inertia and reduced gravitational force. The first term on the right-hand side, $F_z^\mathrm{C}$, is the quasi-steady lift due to the circulation, where $\Gamma \equiv \hat{\Gamma}/(\hat{u}_\mathrm{rel}\hat{l})$, and $u_\mathrm{rel} \equiv (u_\mathrm{b}'^2 + v_\mathrm{b}^2)^{1/2}$. While various models have been proposed to estimate fluid circulation (Tanabe & Kaneko, 1994; Pesavento & Wang, 2004), here $\Gamma$ is obtained from stationary simulations at $Re = 30$ as in Li *et al.* (2022). The second term on the right-hand side is the quasi-steady drag, $F_z^\mathrm{D}$. Here, $C_\mathrm{D}^0$ and $C_\mathrm{D}^{\pi/2}$ denote the drag coefficients of a flat plate at angles of attack $\alpha = 0$ and $\alpha = \pi/2$, respectively. Drag coefficients are defined by dividing the drag force by $\hat{\rho}\hat{u}_\mathrm{rel}^2\hat{l}/2$. As with $\Gamma$, the drag coefficients are determined from steady-state simulations. A curve-fitting approach, following the method of Li *et al.* (2022), is employed to estimate empirical expressions for $\Gamma$, $C_\mathrm{D}^0$, and $C_\mathrm{D}^{\pi/2}$ (figure 7). A third-order polynomial provides a good fit for $\Gamma$, expressed as $\Gamma = -0.485\alpha^3 + 1.853\alpha^2 - 1.718\alpha - 0.009$ for $Re = 30$ and $\alpha$ in radians. Across the range of motions considered, the drag coefficients exhibit a strong dependence on the Reynolds number. The resulting empirical relations are: $C_\mathrm{D}^0 = 7.36Re^{-0.61}$ and $C_\mathrm{D}^{\pi/2} = 8.37Re^{-0.39}$. The third term of equation 10, $F_z^\mathrm{AM}$, is the added mass force, where $m_\mathrm{a} = \pi/4$ (Limacher *et al.*, 2018). Finally, the last term of equation 10, $F_z^\mathrm{UW}$, is the unsteady wake contribution term. This is computed by the difference with the other terms,

$$F_z^\mathrm{UW} = F_z - F_z^\mathrm{C} - F_z^\mathrm{AM} - F_z^\mathrm{D}. \tag{11}$$

The time–history of the force components in equation 10 shows that the vertical force surge in response to the gust is due to circulatory lift for the plate (figure 8). In fact, the circulatory lift exhibits a pronounced peak immediately following gust onset, driving a net surge in the total vertical force. Instead, the vertical component of the drag decreases as the vertical velocity decreases (figure 8c). Therefore, the gust results in the plate gaining horizontal velocity and thus increasing the relative flow velocity and reducing the angle of attack ($\alpha < \pi/2$). In turn, this results in the generation of lift (figures 9a and 9b). Notably, bluff bodies may also experience a reduction in terminal velocity due to flow fluctuations (Fornari *et al.*, 2016). However, in their case, the temporary increase in the upward force is due exclusively to the increased relative velocity and the more-than-linear relationship between drag and relative velocity.

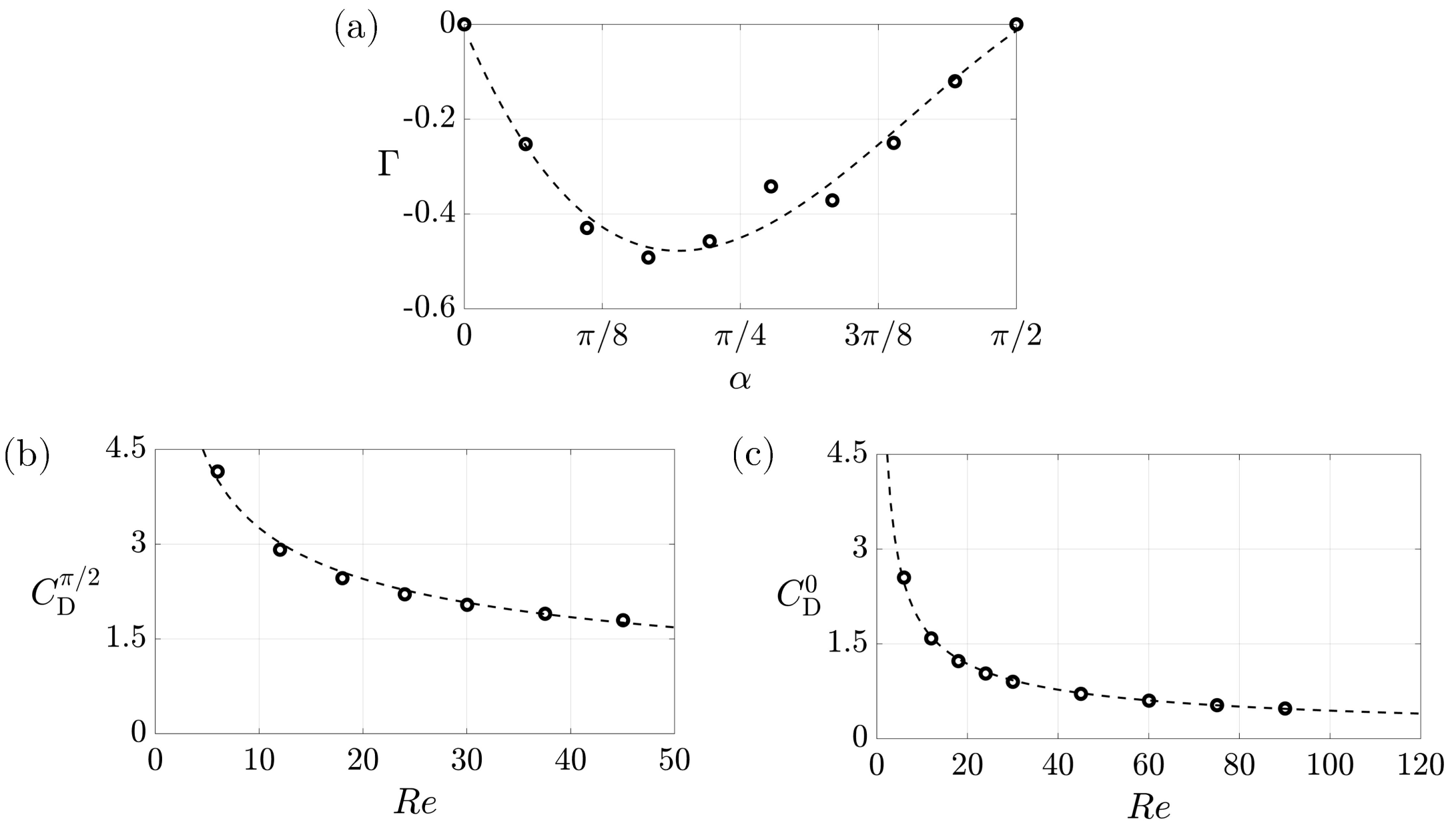

Figure 7: Load coefficients obtained for a stationary plate with $\chi = 10$: (a) circulation versus the angle of attack for $Re = 30$; (b-c) drag coefficient versus Reynolds number for a plate (b) orthogonal and (c) parallel to the stream. Symbols denote computed values, and dotted lines show best fits.

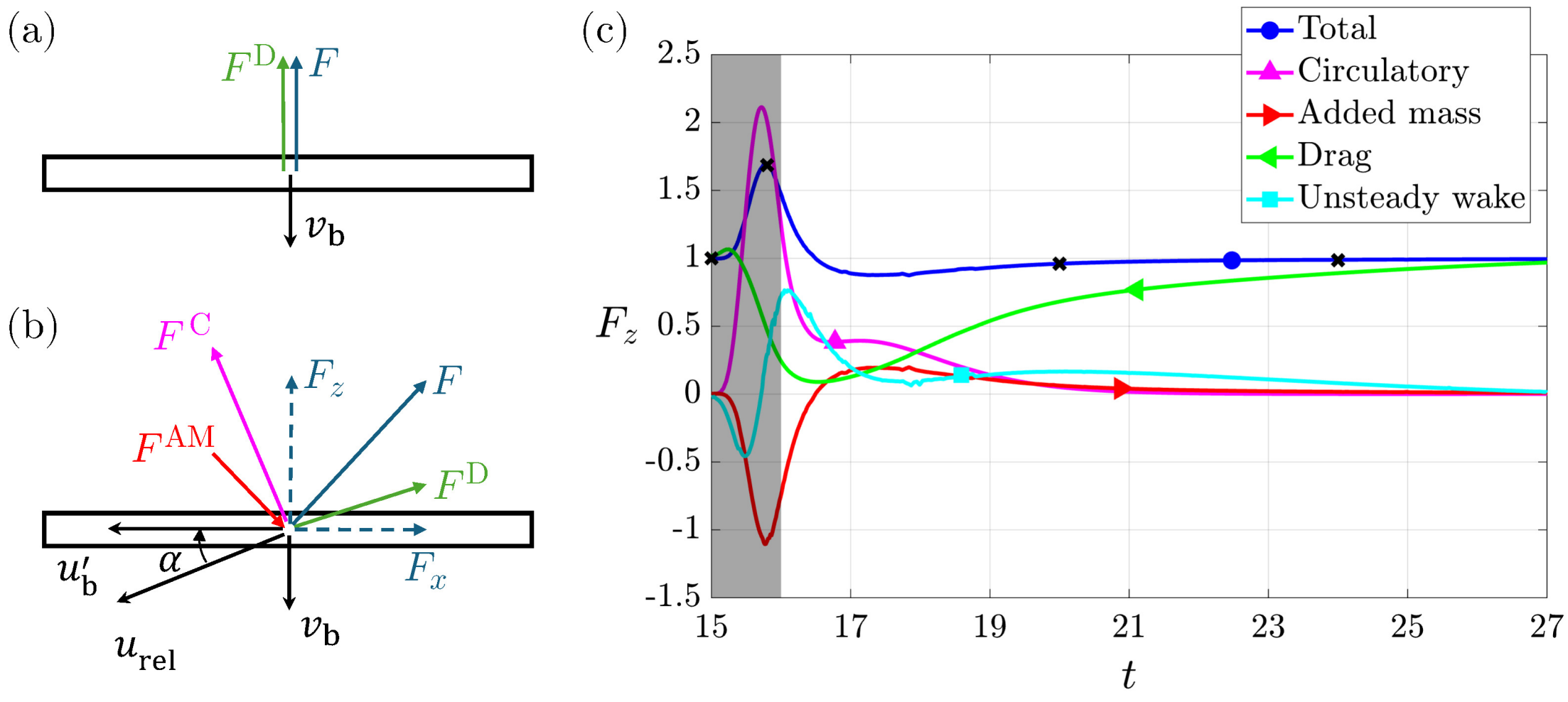

Figure 8: Schematic diagram of the forces on a plate that can translate (two degrees of freedom) but cannot rotate, (a) before the gust and (b) when the vertical force is maximal. (c) Time history of the vertical forces acting on the plate at $Ga = 30$, $\rho = 5$ and $G_R = 5$. The grey-shaded region shows the gust duration between $t = 15$ and $t = 16$.

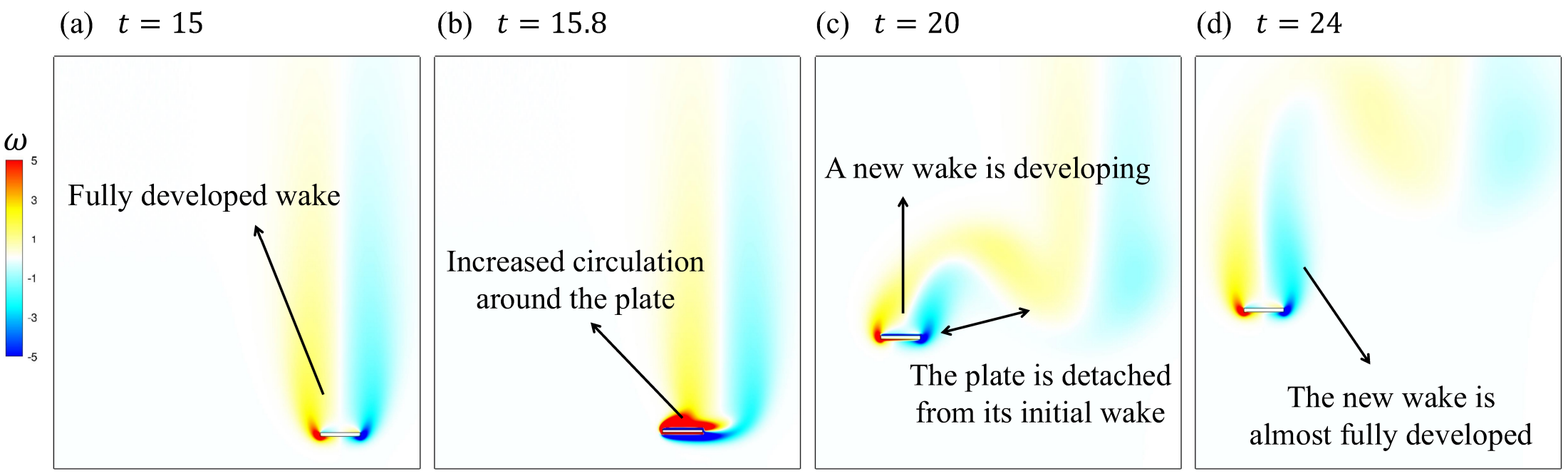

Figure 9: Vorticity contours and description of the most significant phenomena at (a) $t = 15$, (b) 15.8, (c) 20, (d) and 24 for a rotationally-constrained plate with $Ga = 30$, $\rho = 5$, and $G_R = 5$, where the gust begins and ends at $t = 15$ and 16, respectively.

Furthermore, we found a second mechanism that decreases the descent velocity of the plate. The unsteady-wake force $F^{UW}$ remains positive for a long time after the end of the gust. For example, at $t = 30$, while $F^{AM}$ and $F^{C}$ have already vanished, $F^{UW}$ provides a significant contribution to the total vertical force. This effect arises from the temporary detachment of the plate from its wake (figure 9c). Wake detachment causes the plate to behave like an impulsively starting body, generating greater force than it would with a fully developed wake. As the wake returns to its pre-gust state (at $t = 24$ in figure 9d), $F^{UW}$ vanishes and the drag reaches its quasi-steady value, $F^{D}$.

Overall, the uplifting mechanism of a free-falling plate can be summarised as follows. The plate is initially falling steadily at its terminal velocity with a fully developed wake (figure 9a). The plate accelerates horizontally more slowly than the stream because of its greater inertia. Hence, a horizontal velocity is established between the flow and the plate. Combined with the vertical velocity, this results in a greater relative velocity between the plate and the fluid, a reduced angle of attack, and the generation of lift with a positive upward component (figure 9b). The delayed response of the plate to the gust acceleration also results in a horizontal displacement of the plate from its own original wake (figure 9c). Therefore, even after the horizontal velocity with respect to the stream vanishes, its vertical drag is higher than at steady-state because the wake is not yet formed above the plate. Eventually, the plate forms a new wake, and it falls with the same terminal velocity as before the gust (figure 9d).

# 5 Characterisation of the gust response of free-falling plates

In this section, we characterise the gust response of plates with $\chi = 10$, and a uniform mass distribution, i.e. $\mathbf{I}_o = \rho(1 + \chi^{-2})/(12\chi)$. We explore the effect of $Ga \in \{10, 20, ..., 50\}$, $\rho \in \{5, 7, 10, 15, 20, 30, 40, 50\}$, $G_R \in \{1, 2, ..., 5\}$, and gust sharpness $t_G \in \{0.2, 0.5, 1, 2, ..., 5\}$. Unless specifically stated, $t_G = 1$.

We first discuss the effect of the gust on the plate's trajectory, orientation, and altitude gain (§5.1). Subsequently, we explore why the altitude gain is not monotonic with $Ga$ and $G_R$ (§5.2). We achieve this by first demonstrating that the vertical contribution of the pressure force is responsible (§5.3) and then by examining how the pressure force varies during the plate kinematics (§5.4). Finally, we show how the gust response varies with $t_G$ (§C).

## 5.1 Gust effect on plate's trajectory and orientation

Because the transverse gust causes a reduction in the descent velocity (figures 4b and c), at any time after the gust, the plate is at a higher altitude with respect to an earth-fixed frame than it would be without the gust. We found that this for any $Ga$ and $G_R$ (figures 10a-10d). As noted above, for the highest $Ga$ and $G_R$, the vertical velocity can also become temporarily positive (figure 4c), and thus the plate gains altitude compared to its vertical position before the gust (as shown by the increasing altitude $z$ between $t = 16$ and 18 for $G_R = 4$ and 5 in figures 10a-10d).

If released from a fixed height above the ground ($z = 0$ in figures 10), the plate takes longer to reach a lower altitude (e.g. $z = -20$) when it experiences a gust than otherwise. The time during which the plate is airborne is its endurance. We found that for every tested $Ga$ and $G_R$, the gust always increases the endurance of the plate. For example, consider $Ga = 40$ and $G_R = 5$ (figure 10c): the plate reaches altitude $z_b = -15$ at $t \approx 17$ and $t \approx 21$ without and with a gust, respectively. Therefore, the gust increases the plate endurance by $t = 4$.

In general, the endurance of the plate increases with both $G_R$ and $Ga$ (figure 10a and 10b), but the trends are not always monotonic. This is shown by the crossing lines in figure 10c for $G_R = 2$ and 3, and in 10d for $G_R = 2$ and 3, as well as 4 and 5. These non-monotonic trends are related to the plate rotation. In fact, the gust also results in a rotation of the plate (figures 10e-10h). When the terminal velocity is lower than $Re_c$ ($Ga = 20$ and 30 in figures 10e and 10f), the pitch oscillation vanishes over time. Conversely, when the terminal velocity is higher than $Re_c$ ($Ga = 40$ and 50 in figures 10g and 10h), the pitch oscillation becomes periodic.

The plate is horizontal when the pitch angle is $\theta = n\pi$ with $n = 0, 1, 2, \ldots$, and it is vertical when $\theta = \pi(2n - 1)/2$ with $n = 1, 2, 3, \ldots$. The time history of the pitch angle plays a major role in determining the final endurance of the plate and is responsible for the non-monotonic trends with $G_R$ (figures 10c and 10d). In fact, when the plate is vertical, it tends to fall faster than when it is horizontal. Furthermore, a plate that passes swiftly through a vertical orientation to reach a higher pitch loses less altitude than a plate that has a low angular velocity when vertical. In fact, within the set of tested $Ga$ and $G_R$, the non-monotonic trends of endurance with $G_R$ occur because the maximum pitch angle reached by the plate in response to the gust is about $\theta = \pi(2n - 1)/2$. This is because, by definition, the angular velocity is null at the maximum pitch angle. For example, at $Ga = 50$, the maximum pitch angle is about $\pi/2$ for $G_R = 3$ and $3\pi/2$ for $G_R = 5$ (figure 10g). Consequently, the plate at $G_R = 3$ loses altitude compared to that at $G_R = 2$, and the plate at $G_R = 5$ loses altitude compared to that at $G_R = 4$.

## 5.2 Gust effect on plate's altitude gain

The gust effect can also be quantified by the altitude gain $Z_G$, which we define as the difference in altitude $z$ between a plate that has experienced a gust and one that hasn't at a time $t \gg t_s$. The altitude gain varies over time immediately after the gust, but it tends to a constant value for stable plates ($Ga \in \{10, 20, 30\}$), and to a constant mean value for periodically fluttering plates ($Ga \in \{40, 50\}$) (figures 10a-10d). Here, $Z_G$ is evaluated by taking the mean over four periods between $t = 25$ and 30.

The trend of $Z_G$ with $G_R$ for different $Ga$ values clearly shows that the effect of the gust increases non monotonically with $G_R$ (figure 11). In fact, the plates whose maximum pitch angle is about $\theta = \pi(2n - 1)/2$, which are $G_R = 3$ at $Ga = 40$ and $G_R = 3$ and 5 at $Ga = 50$, stand out for being out of trend.

The gust response is also non-monotonic with the density ratio. We showed in section 3 that the density modulates the amplitude of the fluttering oscillations of plates falling in a quiescent flow at $Re_t > Re_c$. Specifically, the maximum fluttering amplitude is observed for $\rho = 20$. A similar trend and maximum are observed for the gust response. Notably, this also occurs for plates whose terminal velocity is such that $Re_t < Re_c$. For example, a plate with $Ga = 30$, falls stably both before and sufficiently after the gust for any value of $\rho$ (figure 12a). Yet, the altitude gain $Z_G$ is maximal for $\rho = 15$ (figure 12b). The underlying mechanism responsible for this behaviour is left for future investigation.

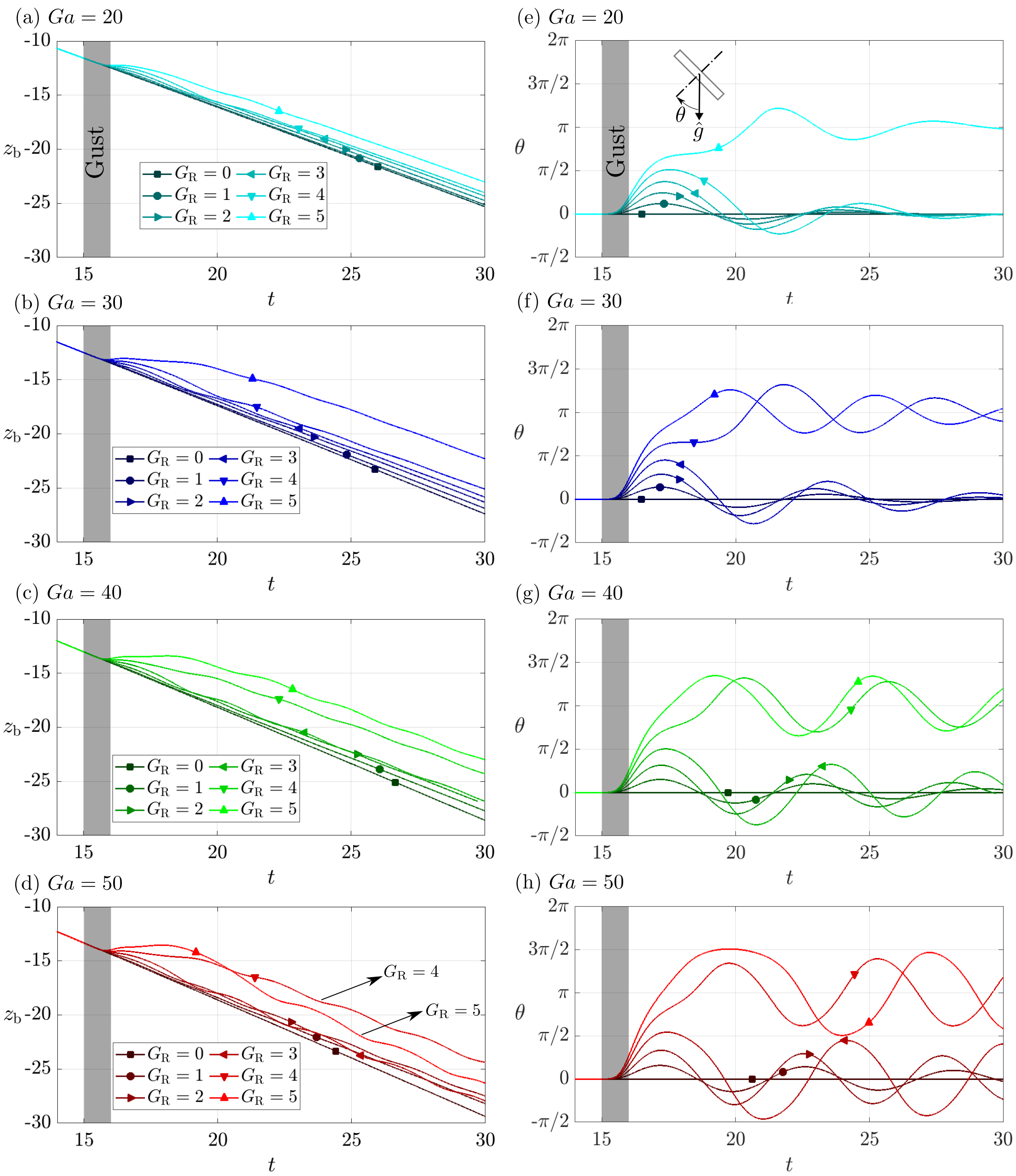

Figure 10: Time history of the plate's (a-d) altitude with respect to an earth-fixed frame, and pitch angle (e-h), for $Ga = 20$ to 50, and $\rho = 5$. The grey-shaded region indicates the gust duration between $t = 15$ and $t = 16$. The inset in (e) depicts the pitch angle of the plate.

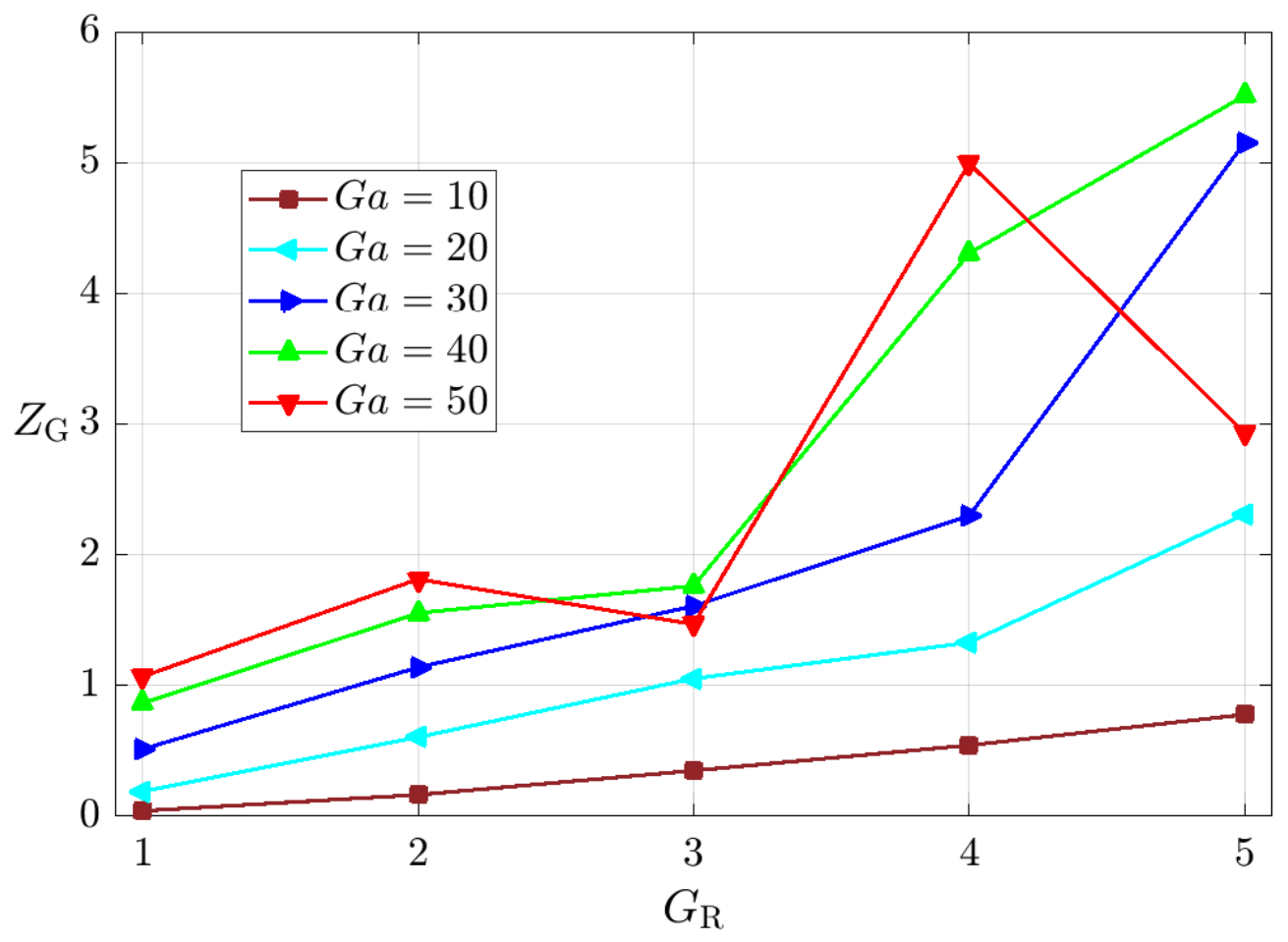

Figure 11: Altitude gain versus the gust ratio for different $Ga$ values and $\rho = 5$.

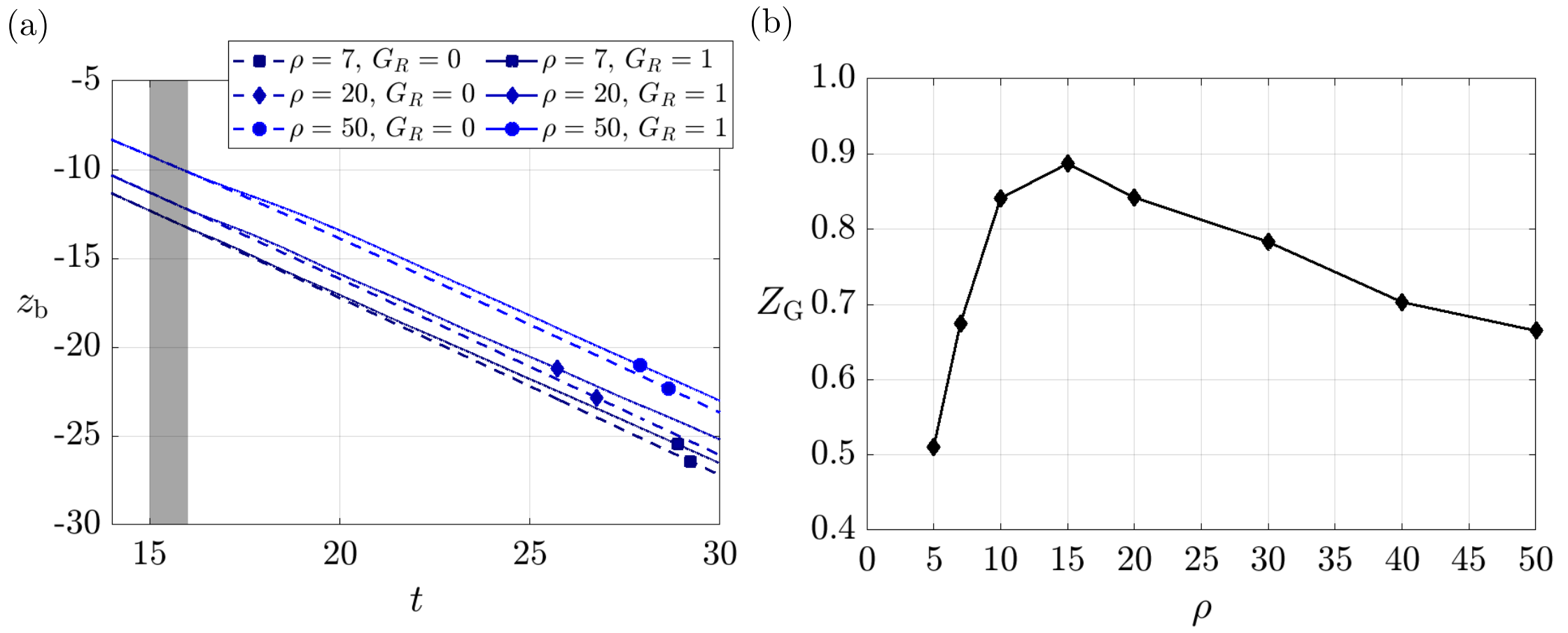

Figure 12: (a) Time history of the plate's altitude with respect to an earth-fixed frame for $Ga = 30$ and different density ratios and gust ratios $G_R = 0$ and 1, and (b) altitude gain versus the density ratio for $Ga = 30$ and $G_R = 1$. The grey-shaded region in (a) shows the gust duration between $t = 15$ and 16.

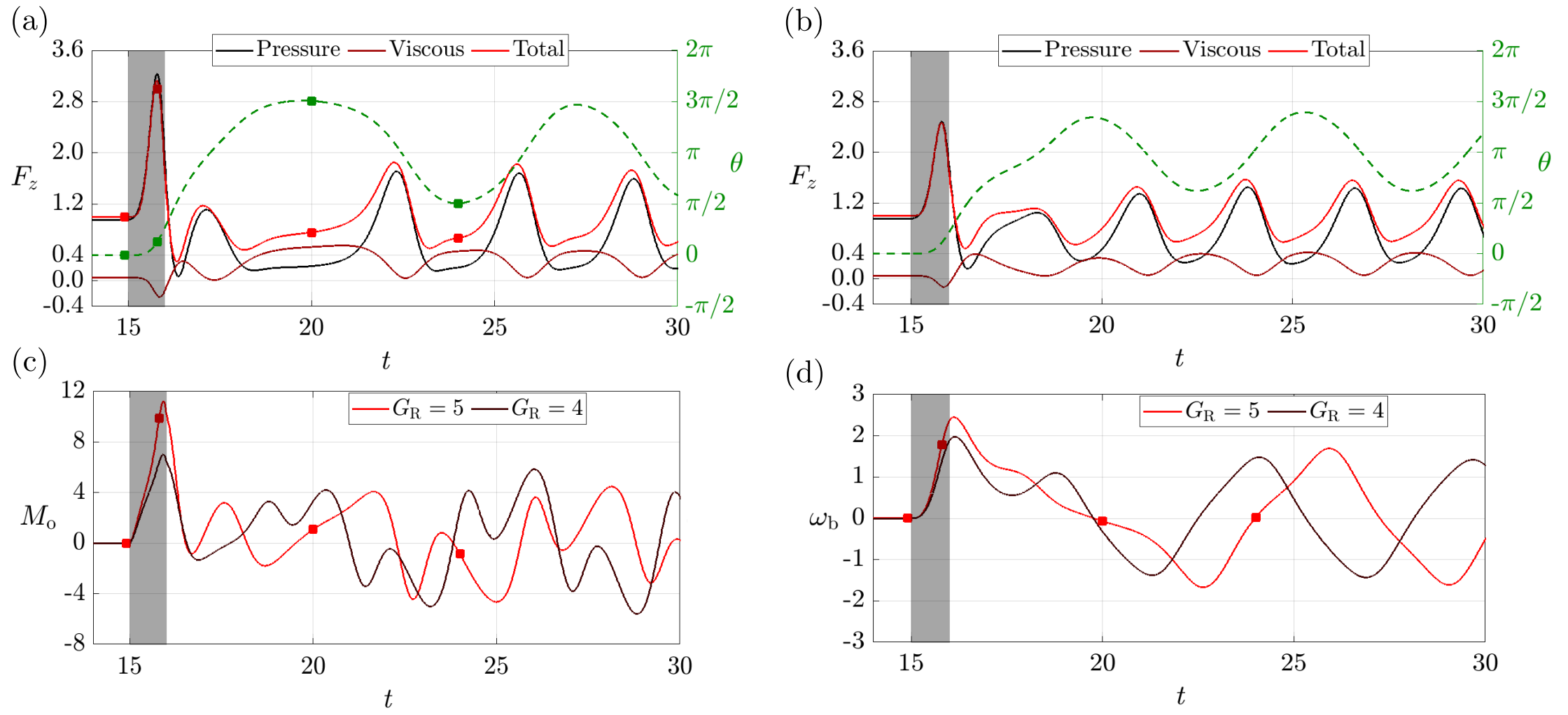

Figure 13: Time history of the vertical force, of its pressure, and viscous contributors, and of the pitch angle for $Ga = 50$, $\rho = 5$, (a) $G_R = 4$ and (b) $G_R = 5$. (c) Pitching moment about the body's centre and (d) body's angular velocity. The grey-shaded region denotes the gust duration between $t_s = 15$ and $t_e = 16$.

### 5.3 Pressure and viscous vertical force components

We showed that the altitude gain is not monotonic with the $G_R$ (figure 11) because, when the maximum pitch is about $\pi(2n-1)/2$, the plate loses altitude (figure 10d and 10h). Therefore, here, we also explore the pitching moment $My$ and the angular velocity $\omega_y$, and how the relative role of pressure and viscous forces underlie this non-monotonic trend. We consider two plates with $Ga = 50$, $\rho = 5$, and $G_R = 4$ and 5. These two plates are selected because the plate with the higher $G_R$ of 5 eventually gains a lower altitude than that at a $G_R$ of 4 (figures 10h), being one of the cases with a non-monotonic altitude gain with $G_R = 4$ (11).

For both $G_R = 4$ and 5, the total vertical force increases as soon as the gust begins ($t = 15$) and shows a peak immediately after the gust (figure 13a and 13b, respectively). By breaking down the vertical force in the pressure and viscous contributions, we observe that this initial force peak is due to the pressure force (figure 13). In fact, the viscous force initially decreases in response to the gust. This is consistent with the increase in the vertical force contribution due to circulation and the reduction of that due to drag shown in section 4.4.

The pressure force considered here is the net of the hydrostatic pressure forces, resulting in buoyancy forces caused by gravitational and gust acceleration, respectively. The total loads experienced by the plate includes the fluid force ($\boldsymbol{F}$) and torques ($\boldsymbol{M}_o$) due to the pressure $p'$ and the viscous stresses, as well as the reduced gravity force ($\boldsymbol{\gamma}$) and the reduced gust acceleration ($-m\dot{\boldsymbol{u}}_G^*$).

The two plates also experience a peak in the pitch moment $My$ immediately after the gust (figure 13c), which makes the plates rotate (figure 13d). The time history of the vertical pressure and viscous forces shows that the vertical pressure force is generally higher when the plate is at $n\pi$ and lower when it is at $\pi(2n-1)/2$. Conversely, the vertical viscous force is, in general, higher when the plate is at $\pi(2n-1)/2$ and lower when at $n\pi$ (figure 13a and 13b). This is because the pressure difference across the plate is maximal when the longest edge of the plate is stream-normal and minimal when it is perpendicular to the stream. Conversely, the vertical shear force is maximal when the longest edge of the plate is vertical, and minimal when it is horizontal. Therefore, the oscillations of the vertical pressure and viscous forces are modulated by the pitch angle.

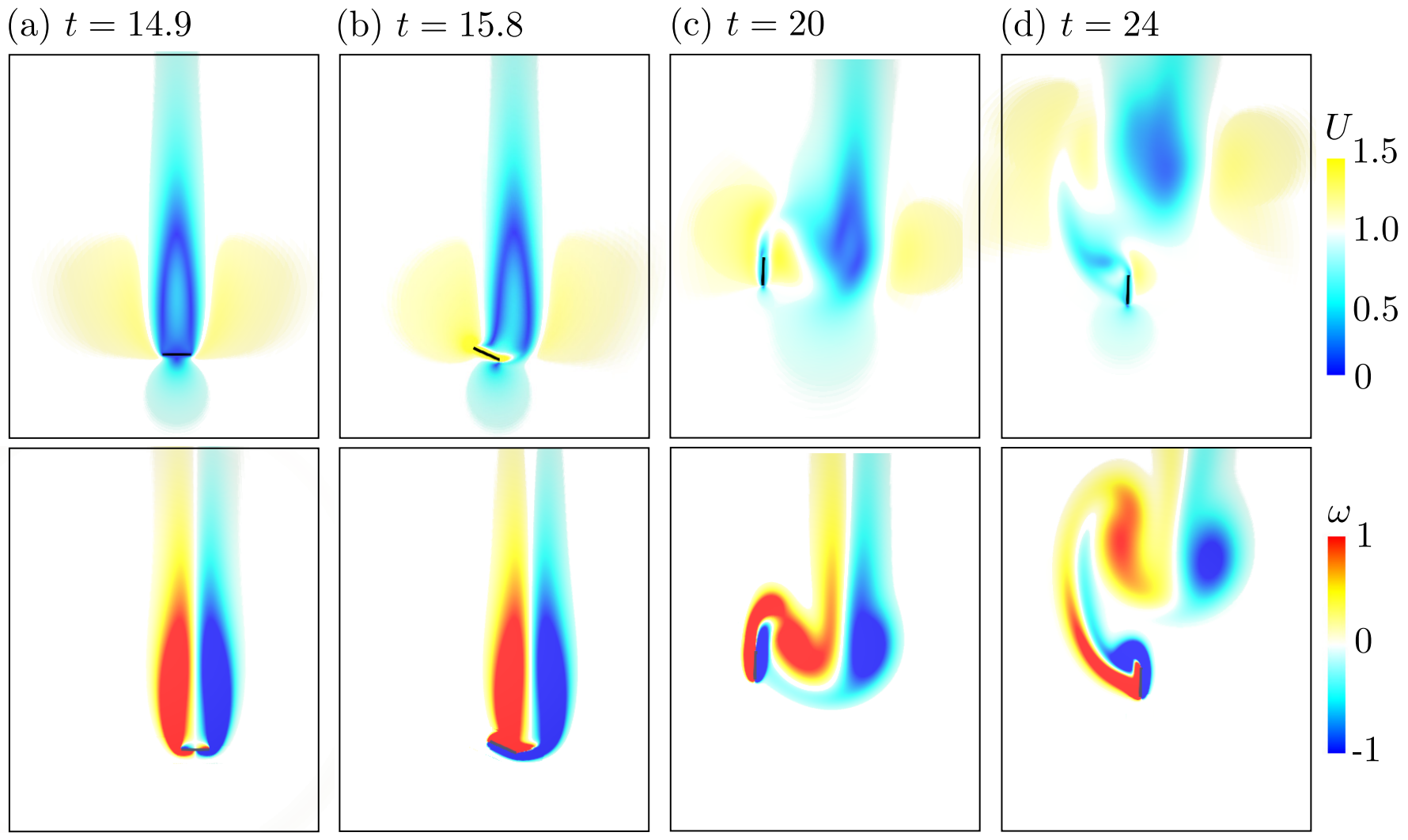

Figure 14: Instantaneous velocity ($U \equiv (u'^2 + v^2)^{1/2}$) and vorticity fields around the plate. Results are presented for a plate at $Ga = 50$, $\rho = 5$, and $G_R = 5$, at (a) $t = 14.9$; (b) $t = 15.8$; (c) $t = 20$; (d) $t = 24$. The time instants (a)-(d) correspond to the solid markers in figure 13.

The plate at $G_R = 4$ and 5 reaches the maximum pitch at $t \approx 20$ (figure 10h), when the angular velocity $\omega_y$ changes sign (figure 13d). For $G_R = 5$, the pitch angle at this time is about $3\pi/2$ and the vertical pressure force is almost at its minimum (figure 13b). Consequently, approaching a maximum pitch angle of $3\pi/2$, the plate loses vertical pressure force and, thus, some of the altitude that it had initially gained at the beginning of the gust (figure 10h). This results in the final lower altitude gain $Z_G$ of the plate at $G_R = 5$ than 4 (figure 11).

### 5.4 Flow field

In this section, we present the velocity and vorticity fields at $Ga = 50$, $\rho = 5$, and $G_R = 5$. The leading edge suction (figure 14a) moves the centre of pressure towards the leading edge, causing the plate to pitch up (figure 14b). After the gust ends ($t > 16$), the plate tumbles and reaches a maximum pitch angle $\theta = 3\pi/2$ at $t = 20$, such that the plate is vertical (figure 13c). Subsequently, the pitch angle decreases and, at $t = 24$, the plate is at $\theta = \pi/2$.

## 6 Concluding Remarks

We investigated, using computational fluid dynamics simulations, the dynamic response of free-falling plates to transverse gusts. We considered two-dimensional plates with an aspect ratio of ten, free-falling under gravity in a quiescent flow. After having been dropped from a stationary position and having reached the terminal velocity, they experience a transverse gust. The gust is a spatially uniform acceleration of the free stream velocity in a direction orthogonal to that of gravity. The strength of the gust is given by the gust ratio, which is the ratio between the free stream velocity increment and the estimate of the terminal velocity of the body. The gust occurs over approximately one convective period, and we show that shorter gust durations would give similar results. We tested plates with Galilei numbers from 10 to 50, density ratios from 5 to 50, and gust ratios from 0 to 5.

We discovered that the gust always results in a temporary reduction in the plate's descent velocity. Therefore, the gust increases the endurance of plates for all values in the tested parameter space. For high Galilei numbers and gust ratios, the plate also reaches higher altitudes than when it experienced the gust. We demonstrated that this uplifting mechanism arises from two phenomena. First, the gust is initially experienced by a vertically-falling plate as a horizontal force. This force displaces the plate horizontally with respect to the free stream velocity, resulting in a temporary reduction in the angle of attack to a lower than $\pi$ value. This allows the plate to generate lift, whose vertical component slows the plate's descent velocity. Second, as the plate is horizontally displaced from its original wake, it temporarily experiences a greater vertical drag component until a fully developed wake is restored.

In general, the altitude gain increases with the Galilei number and the gust ratio. The plate's rotation in response to the gust can further increase the altitude gain. However, if the maximum pitch angle of the plate is about vertical, the plate loses altitude at a high rate until it reaches a horizontal position. This phenomenon results in non-monotonic trends of the final altitude gain with the Galilei number and the gust ratio. Furthermore, rotationally constrained plates develop higher horizontal velocities and reach higher horizontal ranges, resulting in a more persistent uplifting effect than plates free to rotate. Finally, we found that the altitude gain is maximal at a density ratio of about 15. The reason for this local maximum will be the objective of future investigations.

Overall, these results reveal that the effect of a free-falling plate experiencing a transverse gust is that of a temporary decrease in its descent velocity. While the uplifting mechanism has been observed across all tested cases, further work is needed to determine whether there are conditions (including tridimensional bodies) under which gusts increase the descent velocity of free-falling bodies. This work is limited to Galilei numbers sufficiently high to ensure that the gravity force dominates viscous forces, but sufficiently low to ensure that the plate falls steadily before the gust. As the gust response increases with Galilei numbers, future work should explore the effect of gusts on plates that are initially falling with different unsteady kinematics. Finally, transverse gusts are a simplified model of one of the disturbances that a body experiences when free-falling in turbulence. To this end, further work is needed to demonstrate whether the observed underlying mechanisms occur in bodies free-falling in turbulence, in addition to established mechanisms such as non-linear drag, preferential sweeping, and the loitering effect.

These results will directly inform the design of sensors that are passively transported and distributed by the wind. This is an emerging technology with applications spanning environmental monitoring to defence and surveillance. The discovered mechanism enables the flyer to harvest energy from wind fluctuations, thereby increasing its endurance - an effect that can be maximised through design. The gust uplifting mechanism should also be considered to estimate how plant diaspores, debris, etc., are transported by the wind.

**Funding.** This work was supported by the European Research Council through the Consolidator Grant 2020 "Dandidrone: a Dandelion-Inspired Drone for Swarm Sensing" [H2020 ERC-2020-COG 101001499].

**Acknowledgements.** Most of the simulations were carried out at the Cirrus UK National Tier-2 HPC Service at EPCC, Edinburgh.

**Deceleration of Interests.** The authors report no conflict of interest.

Table 1: Numerical error and associated uncertainty due to grid resolution and time step size for $Z_G$ at $Ga = 50$, $\rho = 5$, and $G_R = 5$.

| Case | $h_{grid}$ | $Z_G$ | $\delta_{Z_G}$ (%) | $U_{Z_G}$ (%) | Case | $h_{time}$ | $Z_G$ | $\delta_{Z_G}$ (%) | $U_{Z_G}$ (%) |
|---|---|---|---|---|---|---|---|---|---|
| G1 | 2 | 2.7141 | 13.74 | 17.61 | T1 | 2 | 2.8264 | 6.28 | 7.87 |
| G2 | $\sqrt{2}$ | 2.8696 | 7.79 | 10.16 | T2 | $\sqrt{2}$ | 2.9406 | 2.18 | 2.75 |
| G3 | 1 | 2.9833 | 3.82 | 5.21 | T3 | 1 | 2.9833 | 0.73 | 0.94 |
| G4 | $1/\sqrt{2}$ | 3.0253 | 2.44 | 3.47 | T4 | $1/\sqrt{2}$ | 2.9969 | 0.27 | 0.37 |

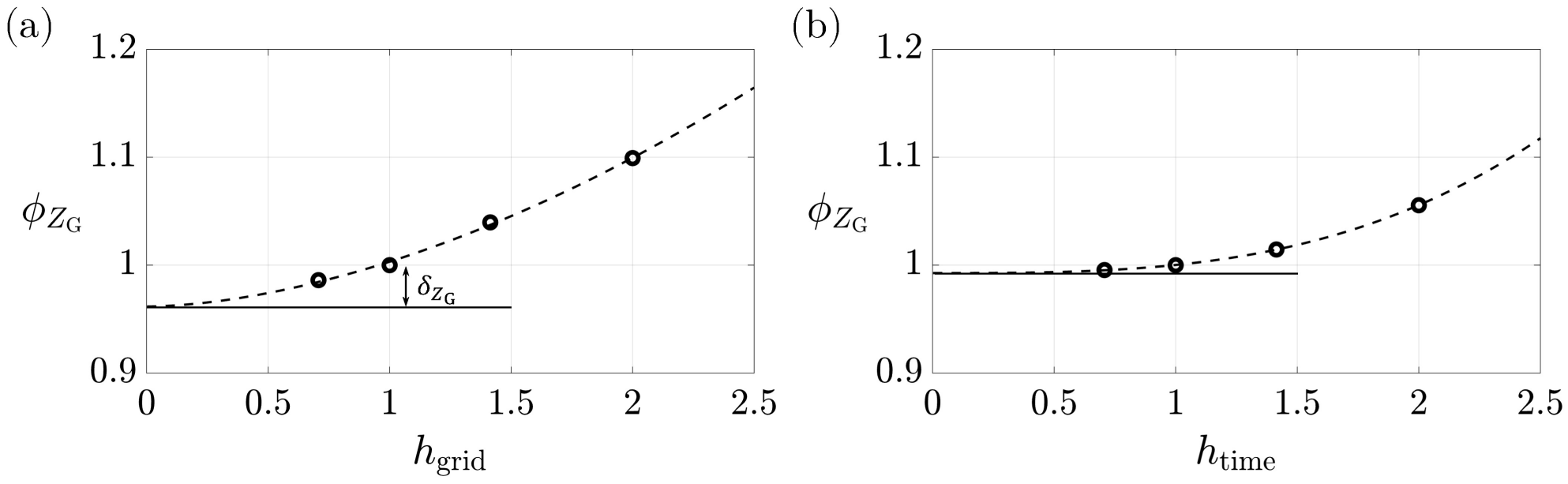

Figure 15: Convergence study for the altitude gain $Z_G$ as a function of (a) grid resolution and (b) time step size.

# A Verification

The uncertainty in the estimate of the altitude gain, $Z_G$ is quantified following the methodology given in Viola *et al.* (2013). This methodology was initially designed for sail yacht aerodynamics but has since been applied to various applications, including permeable plates (Bose *et al.*, 2025) and permeable disks (Cummins *et al.*, 2017). The numerical uncertainty is assumed to result primarily from the grid and time resolution. The reference case of a plate with $Ga = 50$, $\rho = 5$, and $G_R = 5$ is considered as it presents the highest $Re$ conditions. A systematic refinement study was performed using four geometrically similar grids (G1-G4) and four time step sizes (T1-T4), each refined by a factor of $\sqrt{2}$, as shown in table 1. The baseline grid, G3, and time step, T3, are used for the remainder of the study. The relative grid resolution and time step size are defined as $h_{grid} = n/n_{base}$, and $h_{time} = Co/Co_{base}$, where $n$ is the number of grid points around the plate, $Co$ is the Courant number, and $n_{base}$ and $Co_{base}$ are the values of $n$ and $Co$ for the simulation with grid G3 and time step T3. The relative change in $Z_G$ is expressed as $\phi_{Z_G} = Z_G/Z_{G_{base}}$. The convergence of $\phi_{Z_G}$ is estimated using an asymptotic curve,

$$\phi_{Z_G} = ch^p + \phi_0, \tag{12}$$

where $h = h_{grid}$ and $h_{grid}$ for the convergence with decreasing relative grid resolution and relative time step size, respectively. Here, $c$, $p$, and $\phi_0$ are constants determined using the least squares method, and $\phi_0$ represents the converged (true) value as $h \to 0$.

The numerical error is then estimated for both the grid and the time step as $\delta_{Z_G} = \phi_{Z_G} - \phi_0$ (figure 15). For $p > 0.9$, the uncertainty at 95% confidence level in the estimate of $Z_G$ is estimated as

$$U_{Z_G} = 1.25\delta_{Z_G} + \sigma, \tag{13}$$

where $\sigma$ is the standard deviation of the least squares method. The resulting uncertainties due to the finite grid resolution and time step size are 5% and 1%, respectively, for the baseline grid and time step. The combined numerical uncertainty is the Euclidean norm of the two uncertainties, approximately 5%.

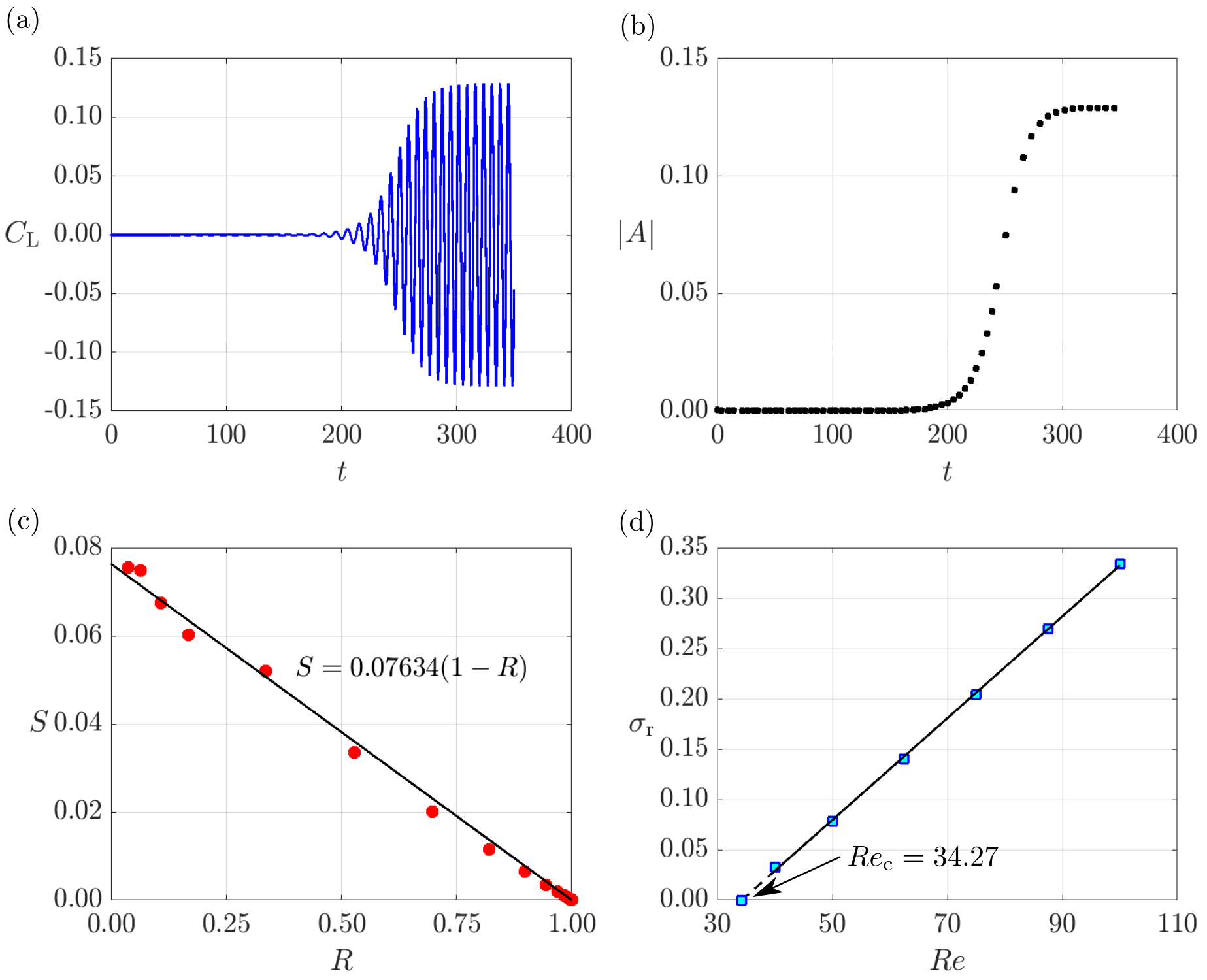

Figure 16: (a) Time history of the lift coefficient, and (b) of the magnitude of its oscillations, and (c) values of $S$ and $R$ at different times for a fixed plate at $Re = 50$, including the linear fit $S = \sigma_r(1 - R)$ giving $\sigma_r = 0.07634$. (d) The slope of the linear fit versus the Reynolds number and the extrapolated critical Reynolds number for $\sigma_r \longrightarrow 0$.

# B Onset of vortex shedding on a fixed plate

The critical Reynolds number is obtained by performing a set of simulations of the fixed plate at Reynolds numbers above the onset threshold, starting each case from a stable wake and allowing the flow to evolve until periodic vortex shedding develops. From the time-history of the lift-coefficient, $C_L \equiv 2F_x$ (figure 16a), the amplitude of successive oscillations, $|A|$, is extracted (figure 16b) and used to compute the instantaneous growth rate $S$ and the normalised squared amplitude $R$,

$$S \equiv \frac{1}{|A|}\frac{\mathrm{d}|A|}{\mathrm{d}t}, \tag{14}$$

$$R \equiv \frac{|A|^2}{|A_{sat}|^2}, \tag{15}$$

where $|A_{sat}|$ is the saturated amplitude once shedding is fully established. These quantities are then fitted with the Stuart–Landau relation $S = \sigma_r(1 - R)$, so that the linear fit yields the instability growth rate $\sigma_r$ for each Reynolds number (figure 16c). Repeating the procedure over several Reynolds numbers gives $\sigma_r(Re)$ (figure 16d). The critical $Re$ is finally identified by extrapolating this curve to the neutral-stability condition $\sigma_r \to 0$, which is $Re_c = 34.27$. This result is consistent with previous studies on plates with $\chi$ between 4 and 10, which suggest that the wake of a plate becomes unstable at $Re$ between 30 and 34 (Saha, 2007; Mashhadi *et al.*, 2021; Rastan *et al.*, 2022).

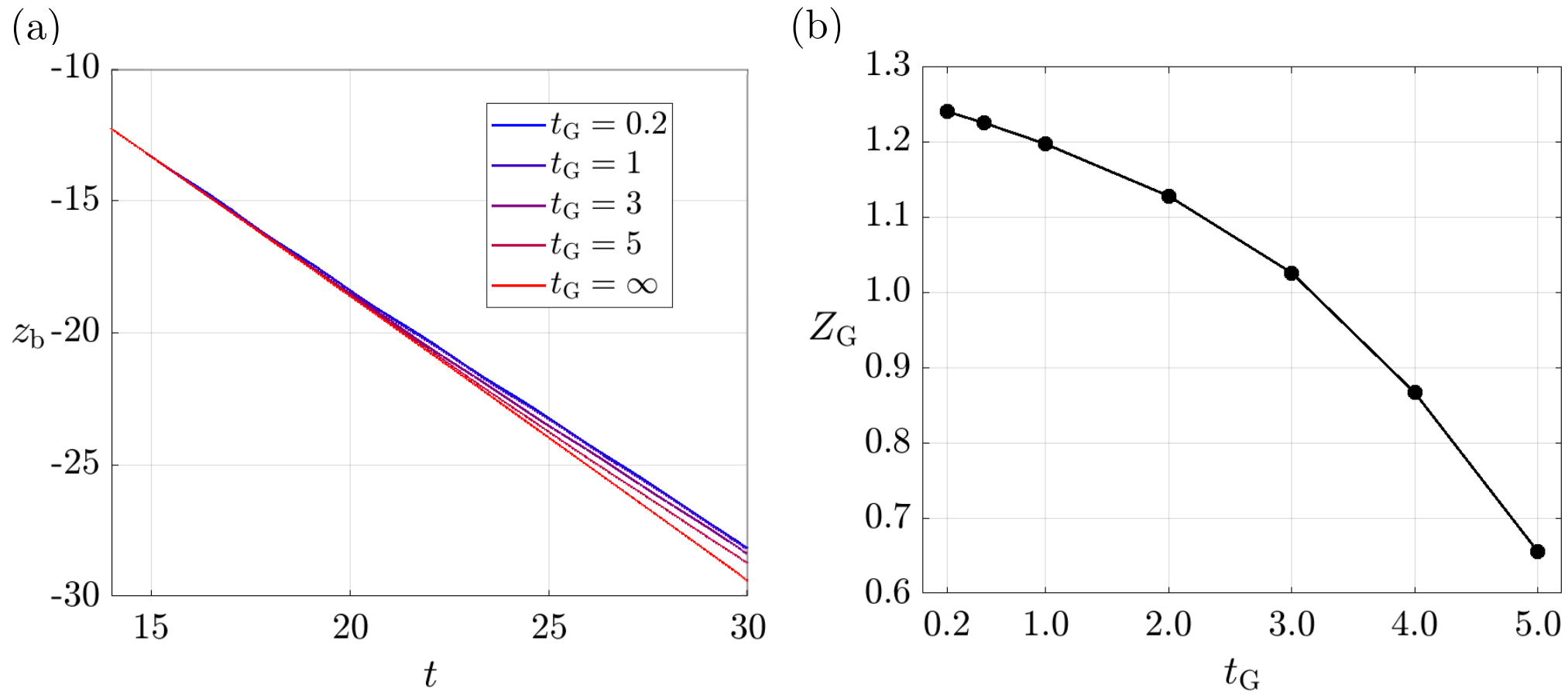

Figure 17: (a) Time history of the plate's altitude with respect to an earth-fixed frame for different gust durations, and (b) altitude gain versus the gust duration for $Ga = 50$, $\rho = 5$, and $G_R = 1$. All gusts start at $t = 15$.

# C Effect of the gust sharpness

We showed in section 2.1 that, for a cosine-shaped gust, the time history of the reduced gust acceleration $\dot{u}^*_G$ is uniquely defined by the gust ratio $G_R$ and the gust duration $t_G$. While $G_G$ sets the strength of the gust, $t_G$ sets its sharpness. In the limit of $t_G \to \infty$, the gust is a quasi-steady variation of the stream velocity, and the effect of the gust vanishes. Instead, in the limit of $t_G \to 0$, the gust is an instantaneous step change in the stream velocity from 0 to $(1 - \rho^{-1})^{-1} G_R$.

While the magnitude of the gust force, $|-m\dot{u}^*_G|$, increases with decreasing $t_G$, the duration over which this force is exerted on the plate decreases. Specifically, as $t_G \to 0$, the maximum of the force magnitude tends to infinity with $t_G^{-1}$. However, the impulse, which is the integral of the force over the period that it is exerted, is constant.

Specifically, the impulse of the force is $-m(1 - \rho^{-1})u_G$ (equation 9). The total impulse is the value of the impulse at $t = t_s + t_G$, which is $-mG_R$. This shows that the total impulse is proportional to $G_R$ and independent of $t_G$.

In the limit of $t_G \to 0$, the total impulse is also proportional to the momentum exerted on the plate. Therefore, if the momentum exerted on the plate tends to a finite value for $t_G \to 0$, the gust response must also approach a finite value.

This limiting condition is difficult to resolve numerically; thus, it is examined here by tracking the trend as $t_G$ decreases. In particular, we consider a range of plates with $Ga = 50$, $\rho = 5$, $G_R = 1$, and $t_G$ between 0.2 and 5. The gust starts at $t_s = 15$ for all plates. We found that the altitude gain $Z_G$ increases monotonically with decreasing $t_G$ (figure 17), and tends to a finite value that is less than 5% higher than that for $t_G = 1$. Therefore, while the previous sections (sections 4 and 5) considered $t_G = 1$, these results are expected to stand for sharper gusts.